\documentclass[fleqn,usenatbib]{mnras}

\usepackage{newtxtext,newtxmath}

\usepackage[T1]{fontenc}

\DeclareRobustCommand{\VAN}[3]{#2}
\let\VANthebibliography\thebibliography
\def\thebibliography{\DeclareRobustCommand{\VAN}[3]{##3}\VANthebibliography}

\usepackage{graphicx}
\usepackage{amsmath}
\usepackage[english]{babel}
\usepackage{ulem}
\usepackage{bm}
\usepackage{subcaption}
\usepackage{array}
\usepackage{diagbox}
\usepackage{dsfont}
\usepackage{xspace}

\newcommand{\Bs}{\ensuremath{\beta_{\rm s}}}
\newcommand{\planck}{{\sl Planck}\xspace}

\title[Primordial $B$ modes in the presence of polarised point sources]{Robust paths to primordial $B$ modes in the presence of polarised point source contamination}

\author[Arianna Rizzieri et al.]{
Arianna Rizzieri$^{1}$\thanks{E-mail: arianna.rizzieri@physics.ox.ac.uk},
Giorgia Zagatti$^{2,3}$,
David Alonso$^{1}$
\\
$^{1}$Department of Physics, University of Oxford, Denys Wilkinson Building, Keble Road, Oxford OX1 3RH, United Kingdom
\\
$^{2}$Dipartimento di Fisica e Scienze della Terra, Università degli Studi di Ferrara, via Saragat 1, I-44122 Ferrara, Italy
\\
$^{3}$Institut d'Astrophysique Spatiale, CNRS, Univ. Paris-Sud, Universit\'{e} Paris-Saclay, B\^{a}t. 121, 91405 Orsay cedex, France
}

\date{Accepted XXX. Received YYY; in original form ZZZ}

\pubyear{\the\year{}}

\begin{document}
\label{firstpage}
\pagerange{\pageref{firstpage}--\pageref{lastpage}}
\maketitle

\begin{abstract}
  A detection of primordial $B$ modes in the Cosmic Microwave Background (CMB) polarisation relies on a robust control and removal of contamination from astrophysical polarised foreground emission. We examine how much polarised extragalactic point sources may contaminate the cosmological signal in ground-based CMB experiments, propagating their impact through component separation to the final parameter constraints. We discuss masking and modelling strategies to ensure that the resulting constraints remain both accurate and precise under this source of contamination. We produce synthetic data containing polarised emission from extragalactic radio sources (ERSs) and dusty star-forming galaxies (DSFGs) following state-of-the-art models for both samples. We find that, while the DSFG contribution is likely too small to cause any significant bias to the cosmological or foreground parameters, the ERS contribution instead, if left unmasked, leads to a bias in the parameters describing the Galactic synchrotron and lensing $B$-mode components. For experiments targeting noise levels of order $\sigma_P\simeq 1\,\mu K\,{\rm arcmin}$, we find that masking all ERSs with fluxes above $\sim$0.1~Jy, coupled with a relatively simple model to marginalise over residual point-source contamination, achieves a suitable balance that allows for the robust recovery of unbiased cosmological and foreground parameters at a negligible cost in parameter uncertainties.
\end{abstract}



\section{Introduction}
  Current CMB experiments are mapping the microwave sky with unprecedented sensitivity, with the primary goal of detecting or further constraining the primordial $B$-mode signal. Since primordial $B$-mode polarisation can only be generated by tensor perturbations in the metric, its detection would provide a powerful probe of primordial gravitational waves. The amplitude of the tensor contribution is commonly parametrised by the tensor-to-scalar ratio, $r$, for which the current upper bound is $r<0.032$ at $95\%$ confidence level \citep{tristram2022improved}. Current and upcoming experiments aim to further improve the sensitivity on $r$, targeting $\sigma(r)\lesssim 0.001$ \citep{abril2026simons,litebird2023probing}.

  However, a detection of the primordial $B$-mode CMB polarisation would need to be demonstrated to be robust against the contamination by polarised astrophysical foreground signals. These foregrounds arise mainly due to two emission mechanisms: thermal emission from interstellar dust and synchrotron radiation from charged particles accelerated by astrophysical processes. Both types of emission originate both in our own Galaxy as well as in other galaxies. From an observational perspective, however, the Galactic and extragalactic contaminants appear fundamentally different because of our location within the Milky Way.

  The foregrounds of Galactic origin appear diffuse across the sky: while they have higher amplitude along the Galactic plane, they also extend to higher Galactic latitudes and are present in any region of the sky. In particular, even at higher Galactic latitudes, the polarised foreground emission is expected to be stronger than the CMB primordial $B$-mode signal \citep{krachmalnicoff2016characterization}. For these reasons, foreground removal is an unavoidable and delicate step in the data analysis aiming at detecting the primordial $B$ modes. Considerable efforts have therefore been dedicated to characterising the polarised Galactic foreground contamination \citep[e.g.][]{krachmalnicoff2016characterization,Planck:2018gnk,Krachmalnicoff:2018imw,Watts:2023nux} and a wide range of component separation techniques have been proposed to clean the CMB signal from these polarised foregrounds. Including both techniques that aim at modelling the foreground emissions, leveraging on their spectral dependence, either modelling and performing the separation at map-level \citep{brandt1994separation,eriksen2006cosmic,Stompor:2008sf,2020JCAP...06..006D,rizzieri2026cleaning} or at power spectrum-level \citep{ade2015joint,planck2020planck,Wolz:2023lzb,azzoni2021minimal,Azzoni:2022otq}; as well as methods that do not directly attempt at modelling the foreground signal, for instance leveraging on statistical properties of the different foreground components \citep[e.g.][]{ichiki2018delta,carones2023analysis,Leloup:2023vkb,Morshed:2024fow,Tran:2026vdg}.

  Turning instead to extragalactic polarised foregrounds for CMB observations, two types of sources are relevant: extragalactic radio sources (ERSs), whose polarised emission is dominated by synchrotron radiation \citep{toffolatti1998extragalactic,tucci2011high,planck2016planck,rubino2023quijote}; and dusty star-forming  galaxies (DSFGs), whose polarised emission may arise from thermal radiation by aligned dust grains in the interstellar medium \citep{bonavera2017statistics,shablovinskaia2025alma}. Over the frequency range and angular scales most relevant to the primordial $B$-mode search, the polarised extragalactic signal is expected to be generally subdominant to Galactic foreground emission and manifest in the CMB maps with a clearly different spatial morphology. In particular, the emission associated with a single galaxy is spatially localised and, depending on the angular resolution and sensitivity of the telescope, it may or not be individually detected and resolved. Because these objects are compact relative to the angular resolution of typical CMB observations, they are commonly referred to as point sources (PSs). In principle, there are two contributions to the angular power spectrum of the emission from each population: a Poisson noise term, as the sources are assumed to be randomly and independently distributed in the sky; and a clustering 1-halo term (note that no 2-halo term is expected, as there is no reason to expect correlations among polarisation of galaxies within different haloes).

  The standard treatment of extragalactic sources in CMB data-analyses focuses around two aspects. Sources detected above a given significance or flux threshold are identified, through source-detection algorithms, compiled into catalogues ~\citep[e.g.][]{planck2016planck,meyers2017southern,perrott202128,vargas2026atacama}), and masked from the maps. In the analysis of smaller angular scales, where extragalactic source emission becomes increasingly significant relative to the CMB and the other foreground signals, the residual contribution from sources below the masking threshold is then modelled at the power spectrum level~\citep{dunkley2013atacama,george2015measurement,choi2020atacama,Beringue:2025bur,planck2020planck}. Extending this strategy to larger angular scale polarisation measurements is, however, non-trivial. Direct source detection in polarisation is considerably more challenging than in total intensity, owing to the typically low polarisation fractions of extragalactic sources and the correspondingly low signal-to-noise ratio of individual polarised detections \citep{lopez2009polarization,argueso2011filter,herranz2021bayesian}.  Moreover, the extent to which the spectral power law modelling commonly adopted in small scale analyses remains adequate for primordial $B$-mode searches has not yet been fully established. In this regime, the cosmological signal of interest is extremely faint, and even a subdominant residual extragalactic contribution may become relevant for the inference of cosmological and foreground parameters.

  Several studies have forecasted the number counts, polarisation fractions, and angular power spectra of polarised PSs relevant in CMB experiments ~\citep[e.g.][]{toffolatti1998extragalactic, de2005predictions, tucci2011high, tucci2012impact, puglisi2018forecasting, lagache2020impact}. These studies generally predict larger polarised contamination from ERS than from DSFG. And find that, depending on the observing frequencies, source masking thresholds, and instrumental configuration, the residual ERS power can become comparable to the primordial $B$-mode signal around the recombination peak for sufficiently small values of $r$, while remaining less relevant at the reionisation peak.
  Previous studies have also highlighted the potential impact of extragalactic sources on delensing performance \citep{puglisi2018forecasting,qu2025impact}. 
  These conclusions, however, have largely been obtained by comparing the predicted PS power spectra directly with the primordial $B$-mode power spectrum, or by expressing the contamination in terms of an equivalent tensor-to-scalar ratio. Such comparisons provide a useful indication of the characteristic amplitude of the contamination, but do not capture its propagation through a realistic multi-frequency component separation and cosmological parameter inference pipeline. A robust assessment of this propagation is still lacking, as is the development of concrete strategies for the removal of resolved polarised sources, and for the modelling of residual contamination.

  In this work we revised such forecasts in view of ongoing and future experiments, such as the Simons Observatory \citep[SO,][]{ade2019simons}, and we perform a full component separation procedure and inference to study the impact of the ERS and DSFG contamination in the data analysis and on the various astrophysical and cosmological parameters.

  The paper is organized as follows. In Sect.~\ref{sec:model}, we discuss the model we implement to forecast the number counts and power spectra for the two types of point sources: the general formalism, the masking strategy we adopt, the model for the ERSs and for the DSFGs. In Sect.~\ref{sec:analysis} we discuss the analysis framework, and in particular how we simulate the total sky signal and covariance, and present the component separation framework and parameter inference. In Sect.~\ref{sec:res}, we present our results, focusing notably on the simulated polarised ERS signal (Sec.~\ref{ssec:res.ersfore}) and its impact on the analysis (Sect.~\ref{ssec:res.ers}), as well as the simulated DSFG polarised signal (Sect.~\ref{ssec:res.dsfg}). Finally, we summarise our conclusions in Sect.~\ref{sec:conclusion}.

\section{Point sources forecasting}\label{sec:model}
  \subsection{Number counts and power spectra}\label{ssec:model.nccl}
    Our goal is to have an estimate for the polarised power spectra due to the point sources that we would measure at all the auto and cross frequencies of the instrument, as well as for the number of sources that we would mask. We are hence interested in forecasting two quantities in particular related to the PS contamination: the number of detected point sources that will be masked, and the power spectrum of the residual point source contamination. We derive expressions for both quantities below. In doing so, we must consider three different frequencies:
    \begin{itemize}
      \item The \emph{reference frequency} $\nu_*$ at which we compute the differential number counts $n_t$of point sources of type $t$,
      \begin{equation}
        \frac{dn_t}{dz}\equiv \frac{d^3N_t}{dz\,d\Omega\,dS_{\nu_*}}.  
      \end{equation}(i.e. the angular number density of sources of type $t$ within an interval of specific flux $S_{\nu_*}$ and redshift $z$ at frequency $\nu_*$).
      \item The \emph{masking frequency} $\nu_m$, at which sources are detected and masked. The final point source mask, used at all frequencies, is defined from these sources (see Section \ref{ssec:model.masking}).
      \item The \emph{measurement frequencies} $\nu$, at which we make maps of the CMB which we use to infer constraints on primordial $B$ modes. In all cases explored here the point source mask is defined at one of these frequency channels (i.e. $\nu_m$ is also one of the measurement frequencies), although this need not be the case.
    \end{itemize}

    \paragraph*{Number counts.} The total number of PS per steradian of type $t$ with a flux above a certain limit $S_{\rm lim}$ in the masking frequency band $\nu_m$ may be calculated as
    \begin{align}
      N^t_{\nu_m}(> S_{\mathrm{lim}})
        &=\int_0 ^\infty dS_{\nu_*}\,\int dz\frac{dn_t(S_{\nu_*},z)}{dz}\,
        P(S_{\nu_m} > S_\mathrm{lim} \mid S_{\nu_*}, t, z),
        \label{eq:ncounts_per_pop}
    \end{align}
    where $P(S_{\nu_m}>S_{\rm lim}|S_{\nu_*},t,z)$ is the probability for a source at redshift $z$ and of type $t$ to have a flux at frequency $\nu_m$ above the threshold, given its flux at frequency $\nu_*$. For a source with a spectral energy distribution (SED) determined by a set of parameters $\boldsymbol{\theta}_t$, this probability is simply a Heaviside function (1 when the flux $S_{\nu_m}$ extrapolated from $S_{\nu_*}$ using its SED is larger than $S_{\rm lim}$, and 0 otherwise), which we write as $\Theta(S_{\nu_m}(S_{\nu_*},\boldsymbol{\theta}_t,t,z)>S_{\rm lim})$. Our point source model will assume that different sources within a given type $t$ have spectra characterised by different parameters $\boldsymbol{\theta}_t$, with a distribution $p(\boldsymbol{\theta}_t)$. An example of such parameters is the source's spectral index. The total probability for type $t$ is given by the average over this distribution:
    \begin{equation}\label{eq:pdetect}
      P(S_{\nu_m}>S_{\rm lim}|S_{\nu_*},t,z)=\left\langle \Theta(S_{\nu_m}(S_{\nu_*},\boldsymbol{\theta}_t,t,z)>S_{\rm lim})\right\rangle_{\boldsymbol{\theta}_t}.
    \end{equation}
    The total number counts is then the sum of that of individual populations. In some cases, the differential number counts of a given type $t$ may be parametrised as
    \begin{equation}
      n_t(S_{\nu_*},z)=n(S_{\nu_*},z)\,P_{\rm mix}(t|S_{\nu_*}),
    \end{equation}
    where $n(S_{\nu_*},z)$ is the total differential number counts at redshift $z$, and $P_{\rm mix}(t|S_{\nu_*})$ is the probability for a source of flux $S_{\nu_*}$ to be of type $t$.
      
    \paragraph*{Power spectra.} The polarised auto and cross-frequency power spectra due to the residual PS contamination is generally dominated by the Poisson contribution from these sources, with equal contributions in $E$ and $B$ modes. In the simplest case, the power spectrum auto-correlation at frequency $\nu$ is \citep[see e.g.][]{de1999polarization,trombetti2018average,puglisi2018forecasting,lagache2020impact}:
    \begin{equation}
      C_\ell^{BB,\nu\nu} = \frac{\langle \Pi^2 \rangle}{2} \int dS_\nu\,S_\nu^2\,n(S_\nu) = \frac{\langle \Pi^2 \rangle}{2} C_\ell^{TT,\nu\nu},
    \end{equation}
    where $\Pi$ is the polarisation fraction with squared mean $\langle\Pi^2\rangle$. We assume that $\Pi$ is independent of frequency and of the flux density $S_\nu$. While data suggest that this may not be the case at frequencies $\lesssim 20 \, \mathrm{GHz}$, there is no significant evidence yet for it to be the case at higher frequency \citep{puglisi2018forecasting,dezotti2018exploring,gupta2019fractional,datta2019atacama}. The simple expression above is not sufficiently general for our purposes, as we need to calculate the general expression for cross-frequency correlations, accounting for populations of different types as well as the masking strategy used. Writing explicitly the redshift dependence and accounting for the SEDs of different source populations, as well as the contribution from imperfectly masked sources, the general expression for cross-correlations is:
    \begin{align}\label{eq:xcls_BB}
      C_\ell ^{BB, \nu_1 \nu_2}
      =& \sum_t \frac{\langle \Pi_t ^2 \rangle}{2}
      \int_0 ^\infty d S_{\nu_*} \,
      \int dz\,\frac{dn_t(S_{\nu_*}, z)}{dz}\,\langle f_{\nu_1}S_{\nu_1}\,f_{\nu_2}S_{\nu_2}\rangle_{\boldsymbol{\theta}_t},
    \end{align}
    where $f_\nu(S_\nu,t,\boldsymbol{\theta}_t)$ quantifies the fraction of the flux at frequency $\nu$ for a source of type $t$ with SED characterised by $\boldsymbol{\theta}_t$ that has been missed by our masking criterion. This is described in the next section. Note that the power spectra thus computed correspond to maps provided in units of specific intensity (e.g. Janskys per steradian). We then convert these to thermodynamic temperature units ($\mu K_{\rm CMB}$) using the CMB blackbody spectrum for a temperature $T_{\rm CMB}=2.7255\,K$.
    In addition to the Poisson noise term a clustering term may also be in principle relevant, see Sect.~\ref{sssec:model.DSFG.clustering}. 
    
    \vspace{5pt}
    \noindent Note that the flux limit associated with both of these quantities is not given explicitly as an integration limit. Instead, it is encoded in the masking criterion defining $f_\nu$, in the case of the $C_\ell$s, and in the selection criterion ($S_{\nu_m}>S_{\rm lim}$) in the case of the number counts of masked sources. 

  \subsection{Masking strategy}\label{ssec:model.masking}
    We assume that the point source mask used in the analysis is based on the positions of the sources detected at a frequency $\nu_m$. Once defined, the same mask is then used to analyse maps at all other frequencies, and to derive $B$-mode constraints after component separation. This is an important aspect of $B$-mode analysis in the context of component separation, as using the same mask across frequencies ensures that the same foreground spatial modes are present in all cross-frequency power spectra. Besides the position of the masked sources, the other element defining the mask is the masking radius, defining the disk removed around each source. We parametrise this radius as
    \begin{equation}\label{eq:r_mask}
      R_{\rm mask}=\frac{1}{2}f_{\rm FWHM}\,\theta_{\rm FWHM}(\nu_m),
    \end{equation}
    where $\theta_{\rm FWHM}(\nu_m)$ is the full-width at half-maximum aperture of the instrumental beam at the masking frequency $\nu_m$, and $f_{\rm FWHM}$ is a parameter quantifying the fraction of the FWHM aperture removed by the mask. The strictness of the mask is then defined by two parameters: $S_{\rm lim}$ determines the number of sources to mask, and $f_{\rm FWHM}$ the amount of area removed by masking.

    Assuming a Gaussian source profile, the residual flux of a masked source at frequency $\nu$ is
    \begin{equation}
      S_\nu^{\rm res}=S_\nu\,\exp\left[-\ln 2\,f_{\rm FWHM}^2\frac{\theta_{\rm FWHM}^2(\nu_m)}{\theta^2_{\rm FWHM}(\nu)}\right].
    \end{equation}
    Hence, the residual fraction at frequency $\nu$, $f_\nu$, entering the PS power spectrum (Eq. \ref{eq:xcls_BB}) is given by:
    \begin{equation}
      f_\nu=\left\{
      \begin{array}{cc}
        \exp\left[-\ln{2}\,f_{\rm FWHM}^2\frac{\theta_{\rm FWHM}^2(\nu_m)}{\theta_{\rm FWHM}^2(\nu)}\right] & {\rm if}\,\,S_{\nu_m}>S_{\rm lim}, \\
        1 & {\rm otherwise.} 
      \end{array}
      \right.
    \end{equation}
    This quantity depends implicitly on the SED parameters $\boldsymbol{\theta}_t$ through the relation between $S_\nu$ and $S_{\nu_m}$.

  \subsection{Model for polarised extragalactic radio sources}\label{ssec:model.radio}
    Polarised PS emission from extragalactic radio sources is caused by partially polarised synchrotron radiation sourced by active galactic nuclei (AGN) hosting a central super-massive black hole (SMBH). These objects present narrow collimated jets of material ejected from the SMBH and travelling far from the hosting galaxy. Lobes of material, constantly fed by the jets with new energetic particles, form around the jets and extend up to a few megaparsecs from the central SMBH. The relativistic jets emit Doppler-boosted synchrotron radiation with spectrum $S_\nu \propto \nu^{\alpha}$ which, in the frequency range $1-5 \, \mathrm{GHz}$ is approximately flat, with $\alpha > -0.5$, while the lobes emit in radio with a steeper spectral index $\alpha <  -0.5$. Depending on each source's viewing angle, one of these two contributions dominates: edge-on ERSs are typically dominated by the lobes and classified as steep-spectrum sources, while face-on sources are dominated by the jets and classified as flat-spectrum.
    Additionally, a less abundant sample of ERSs exists exhibiting spectra with a positive tilt, and is often referred to as the ``inverted-spectrum'' population. 
    More details can be found in \cite{de2010radio,jackson1999extragalactic} and references cited therein. 

    For our purposes, we are mostly interested in the spectral properties in the frequency range $20\,{\rm GHz}\lesssim \nu\lesssim 300\,{\rm GHz}$, covered by most CMB experiments. We hence consider three ERS types, depending on the value of their effective spectral index in the frequency range 1-5 GHz, $\alpha^5_1$: ``steep'' ($\alpha^5_1<-0.5$), ``flat'' ($-0.5 \leq \alpha^5 _1 < 0.3 $), and ``inverted'' (if $\alpha^5 _1 \geq 0.3$). 
    
    Because of the scarcity of data on polarised ERSs, and as done in previous analyses \citep[e.g.][]{puglisi2018forecasting,lagache2020impact}, we do not consider models for the abundance of polarised ERS number counts directly. Instead, we use models derived from number counts in intensity, and produce predictions for the residual PS polarised power spectra assuming a given polarisation fraction probability density function $\mathcal{P}(\Pi)$. Similarly, and following the approach of \cite{tucci2011high, lagache2020impact}, due to the wealth of data at low frequencies, we will base our calculations on models of the differential number counts of ERSs at $\nu_*=5\,{\rm GHz}$. The emission at the higher frequency range of interest will then be predicted using redshift-independent models for the radio spectra of the different source populations.

    \subsubsection{Differential number counts}\label{sssec:model.radio.nc}
      We use the model for the number counts of sources from all radio populations (steep$+$flat$+$inverted) at $\nu_*=5\,{\rm GHz}$ described in \cite{toffolatti1998extragalactic}. We extrapolate this model linearly in log-log space towards large fluxes ($S_{\nu_*}>10\,{\rm Jy}$), not covered in the original work. We combine this with a model for the joint number counts of flat and inverted sources, also at $5\,{\rm GHz}$, from \cite{tucci2011high}, described as a broken power law fitted to existing data. The abundance of steep sources is then given by the difference between these two combined populations:
      \begin{equation}
        n_{\rm steep}(\nu_*)=n_{{\rm steep}+{\rm flat}+{\rm inv}}(\nu_*)-n_{{\rm flat}+{\rm inv}}(\nu_*).
      \end{equation}
      The number counts of flat and inverted sources are then calculated by rescaling the flat$+$inverted distribution by the odds of a source being of either type given their spectral index. I.e.:
      \begin{equation}
        n_{{\rm flat}/{\rm inv}}(\nu_*)=n_{{\rm flat}+{\rm inv}}(\nu_*)\,P_{\rm mix}({\rm flat}/{\rm inv}|S_{\nu_*}),
      \end{equation}
      where:
      \begin{eqnarray}
        &P_{\rm mix}({\rm flat}|S_{\nu_*})=P(-0.5\leq\alpha^5_1<0.3|S_{\nu_*}),\\
        &P_{\rm mix}({\rm inv}|S_{\nu_*})=P(\alpha^5_1\geq0.3|S_{\nu_*}).
      \end{eqnarray}

      The distribution of effective spectral indices $\alpha_1^5$ for the different populations is modelled, again following \cite{tucci2011high}, as the sum of two truncated Gaussian distributions, one for the steep and one for the flat+inverted populations, with mean and variance dependent on the flux at 5 GHz $S_{\nu_*}$ (see Table 1 in \cite{tucci2011high} for the peak and sigma of these distributions in four flux ranges).

    \subsubsection{High-frequency spectra}\label{sssec:model.radio.seds}
      To estimate source fluxes at the higher frequencies we use here for masking and analysis, we must extrapolate from the flux at the reference frequency $S_{\nu_*}$ (with $\nu_*=5\,{\rm GHz}$) assuming a SED. The starting point for the three ERS populations considered here is the effective spectral index $\alpha_1^5$ in the low-frequency range (which, as we discussed above, is itself a stochastic spectral parameter). The SED of the different populations is then modelled as follows.

      \paragraph*{Steep-spectrum sources.} The spectrum above 5 GHz is modelled as a power law with a high-frequency spectral index $\alpha_h=\alpha_1^5-\Delta\alpha_{\rm steep}$, where the steepening index $\Delta\alpha_{\rm steep}$ is itself a stochastic variable. Following \cite{tucci2011high}, the distribution of $\Delta\alpha_{\rm steep}$ is Gaussian with mean $0.3$ and standard deviation $0.2$. The flux density at frequency $\nu$ is then
      \begin{equation}
        S_\nu = S_{\nu_*}\left(\frac{\nu}{\nu_*}
        \right)^{\alpha_h}.
      \end{equation}

      \paragraph*{Flat-spectrum sources.} We use the {\tt C0} model described in \cite{tucci2011high}. As in the case of steep sources, the spectrum above $\nu_*=5\,{\rm GHz}$ is described as a power law with a steepened spectral index $\alpha_h=\alpha_1^5-\Delta\alpha_{\rm flat}$. The distribution of steepening indices is then modelled as a Gaussian with mean and standard deviations depending on the value of $\alpha_1^5$:
      \begin{equation}
        \Delta\alpha^{\rm flat}\sim
        \begin{cases}
        \mathcal{N}\!\left(0.1,\,0.1\right),
        & -0.5 \leq \alpha_1^5 < -0.1, \\[4pt]
        \mathcal{N}\!\left(0.2,\,0.2\right),
        & -0.1 \leq \alpha_1^5 < 0.3.
        \end{cases}
      \end{equation}
      More complex models for the high-frequency SED of the flat-spectrum population were also proposed by \cite{tucci2011high}. However, we adopt the simpler {\tt C0} as it represents the most conservative scenario, predicting the largest number of sources at high frequencies.

      \paragraph*{Inverted-spectrum sources.} In this case, following \cite{tucci2011high}, the high-frequency spectrum is modelled as a peaked spectrum. The spectrum rises up to a peak frequency $\nu_{\rm peak}$, and then falls with a post-peak spectral index:
      \begin{equation}
        S_\nu=
        \begin{cases}
          &S_{\nu_*}
          \left(\frac{\nu}{\nu_*}\right)^{\alpha_1^5(S_{\nu_*})},
          \qquad \nu \leq \nu_{\rm peak} \\
          &S_{\nu_{\rm peak}}
          \left(\frac{\nu}{\nu_{\rm peak}}\right)^{\alpha_0-\Delta\alpha_{\rm inv}},
          \qquad\nu > \nu_{\rm peak}
        \end{cases}
      \end{equation}
      with $\alpha_0=-0.1$. In this case, both $\nu_{\rm peak}$ and $\Delta\alpha_{\rm inv}$ are stochastic spectral parameters, with the distributions described in \cite{tucci2011high}. $\nu_{\rm peak}$ follows a piecewise-constant distribution (see Table 3 in \cite{tucci2011high}), while the distribution of $\Delta\alpha_{\rm inv}$ is parametrised as:
      \begin{equation}
        p(\Delta\alpha_{\rm inv})\propto\Delta\alpha_{\rm inv}
        \exp\left[-\dfrac{\Delta\alpha_{\rm inv}^2}{2\sigma_\alpha^2}\right],
      \end{equation}
      with $\sigma_\alpha=0.53$, and $\Delta\alpha_{\rm inv}$ constrained to the range $0\leq\Delta\alpha_{\rm inv}\leq2.1$. The distribution of $\alpha_h=\alpha_0-\Delta\alpha$ peaks at $\alpha_h\simeq-0.7$, and can be seen in Fig. 9 of \cite{tucci2011high}.

      \vspace{5pt}
      \noindent To summarise, the spectral parameters for the three populations studied here are:
      \begin{eqnarray}
        \boldsymbol{\theta}_{\rm steep/flat}=\left(\alpha_1^5,\,\Delta\alpha_{\rm steep/flat}\right);
        &\boldsymbol{\theta}_{\rm inv}=\left(\alpha_1^5,\,\nu_{\rm peak},\,\Delta\alpha_{\rm inv}\right).
      \end{eqnarray}
      As we have discussed, these spectral parameters are stochastic, and described by probability distributions which, moreover, are conditional on the population type ($t$) and, in some cases, on the flux at 5~GHz ($S_{\nu_*}$), $p(\boldsymbol{\theta}_t | S_5, t)$. The expectation values introduced in equations \ref{eq:pdetect} and \ref{eq:xcls_BB} are then given by:
      \begin{equation}
        \langle Q (S_{\nu_*}, \boldsymbol{\theta}_t, t) \rangle_{{\boldsymbol{\theta}}_t} = \int d \boldsymbol{\theta}_t\,p(\boldsymbol{\theta}_t | S_5, t) \, Q(S_5, \boldsymbol{\theta}_t, t),
      \end{equation}
      where $Q$ is either $\Theta$ or $f_{\nu_1} S_{\nu_1} f_{\nu_2} S_{\nu_2}$.
      Note that these quantities are independent from the redshift, hence for ERSs equations~\ref{eq:ncounts_per_pop} and \ref{eq:xcls_BB} simplify by taking the expectation values outside the integral over redshift, which allows as to use the differential number counts already integrated over redshift, as presented in Sect.~\ref{sssec:model.radio.nc}.
      We compute these expectation values by deterministic numerical quadrature on a multidimensional grid in the spectral-parameter space. For each spectral parameter $\theta^j$, we define a grid $\{\theta_i^j\}$ and the corresponding normalised quadrature weights
      \begin{equation}
        w_i^j\propto p(\theta_i^j \mid S_{\nu_*},t)\,\Delta\theta_i^j,
        \qquad\sum_i w_i^j = 1,
      \end{equation}
      where $j$ denotes the spectral parameter type and $i$ the grid point. For parameters whose probability distribution does not depend on $S_{\nu_*}$, such as $\Delta\alpha_{\rm steep}$, the grid and its weights can be precomputed. For the others, such as $\alpha_1^5$, the corresponding quadrature weights are instead evaluated at each value of $S_{\nu_*}$. Finally, when the distribution of one spectral parameter depends on another spectral parameter (e.g. $\Delta\alpha_{\rm flat}$ and $\alpha_1^5$), its quadrature weights are evaluated conditionally at each grid point of the latter. Models with a higher-dimensional parameter space are therefore more computationally expensive to calculate.

    \subsubsection{Summary}\label{sssec:model.radio.summary}
      To summarise, the number counts of masked ERSs and the polarisation cross-frequency power spectra of the residual point source contamination after masking are calculated using equations \ref{eq:ncounts_per_pop} and \ref{eq:xcls_BB}. The masking strategy, determining the unmasked fractions $f_\nu$ in Eq. \ref{eq:xcls_BB}, is described in Section \ref{ssec:model.masking}. The differential number counts for the three different source types considered here (steep, flat, and inverted) are described in Section \ref{sssec:model.radio.nc}, while the source SEDs, and the probability distributions of their spectral parameters (required to estimate the ensemble averages in equations \ref{eq:pdetect} and \ref{eq:xcls_BB}) are described in Section \ref{sssec:model.radio.seds}.
      We compute only the Poisson term of the ERS power spectrum as the clustering term is negligible compared to the former \citep{toffolatti1998extragalactic,Gonzalez-Nuevo:2004jlc} unless aggressive flux cuts, $S_\mathrm{lim}\ll10\,\mathrm{mJy}$, are considered, which is not the case in our analysis.
      
      The final ingredient required to predict the level of residual point source contamination from polarised ERSs is the squared-mean polarisation fraction $\langle\Pi^2\rangle$. Here we follow the approach of \cite{lagache2020impact}, assuming a constant polarisation fraction across all sources with $\langle \Pi^2 \rangle^{1/2} =\langle \Pi\rangle=2.8\%$, in agreement with Planck, SPT and ACT data \citep{puglisi2018forecasting,gupta2019fractional,datta2019atacama}. In Section \ref{sssec:res.ers.polfrac} we will consider the impact of scatter in the distribution of polarisation fractions, and instead use a larger value $\langle \Pi^2 \rangle^{1/2} = 4.94\%$ \citep[see][]{puglisi2018forecasting} for the flat-spectrum population, which dominates the residual point source contamination.

      The software developed to compute the number counts of masked ERSs, the polarisation cross-frequency power spectra of the residual contamination after masking and the loss of sky area due to PS masking is made publicly available\footnote{\url{https://github.com/a-rizzieri/ERS_forecast}}.

  \subsection{Model for polarised dusty star-forming galaxies sources}\label{ssec:model.DSFG}
    Dust grains in the interstellar medium absorb optical and ultra-violet radiation and emit radiatively in the far infrared and microwave spectrum with what is typically well described by a modified black body spectrum. Because of their elongated shape, dust grains may partially align with the interstellar magnetic field, and thermal dust emission in dusty star-forming galaxies is therefore expected to show a certain degree of polarisation. The signal we receive from one DSFG, however, is integrated over the whole galaxy, so its observed fractional polarisation is expected to be much lower than that coming from any one point in the galaxy. Few direct measurements exist: for individual objects,~\cite{greaves2002submillimetre} measured a global polarisation degree of only $0.4\%$ at  $353$~GHz for \texttt{M82}, while~\cite{seiffert2007upper} set a $99\%$ confidence upper limit of $1.54\%$ for \texttt{Arp 220} at the same wavelength. Stacking $\sim4700$ DSFGs in the \planck maps,~\cite{bonavera2017statistics} reported mean fractional polarisations of $3.10\pm0.75\%$ and $3.65\pm0.66\%$ at 217 and 353~GHz, respectively. A re-analysis of the same data conducted by~\cite{trombetti2018average} showed no significant detection and derived $90\%$ confidence upper limits of $\langle\Pi\rangle\lesssim2.2\%$ at 353~GHz and $\lesssim3.9\%$ at 217~GHz. On the theoretical side,~\cite{dezotti2018exploring} estimated a mean polarisation degree of $1.4\%$ for spiral galaxies. We hence adopt this latter value and, as in the case of ERSs, we do not model polarised number counts directly, but work with the intensity counts and convert predictions for the residual cross-spectrum in intensity to polarisation through this polarisation fraction.

    In intensity, the spectrum of a DSFG is dominated by a modified black body spectrum, peaking in the far infrared ($T\sim 20\,{\rm K}$) and falling as $\propto\nu^{2+\beta_{\rm dust}}$ towards the millimetre, with $\beta_{\rm dust}$ the dust emissivity spectral index. The frequency range of a typical CMB experiment ($20\,{\rm GHz}\lesssim \nu\lesssim 300\,{\rm GHz}$) samples this steeply rising part of the spectrum, where the DSFG contribution grows rapidly with frequency and may become relevant only towards the top of the band range. 

    We follow the two-population model for the Cosmic Infrared Background (CIB) of~\cite{zagatti2024halo}, which we summarise in here, and extend to polarised point sources. The model is built on the galaxy evolution framework of~\cite{granato2004physical} and~\cite{cai2013hybrid}, a hybrid approach in which massive proto-spheroidal galaxies are treated with a physical, semi-analytic prescription, while late-type and starburst galaxies are described through an empirical approach. Within this framework, DSFGs are split in two samples which occupy dark matter haloes differently and emit at different epochs:
    \begin{itemize}
      \item the early-type (ET) galaxies, are the massive proto-spheroids, which dominate the emissivity at high redshifts, $z\gtrsim1.5$, and reside in the more massive haloes;
      \item the late-type (LT) galaxies, are modelled as the sum of the \textit{cold} (normal late-type) and \textit{warm} (starburst) components, dominate at smaller redshifts, $z\lesssim1.5$, and populate less massive haloes.
   \end{itemize}
   The two populations entering the analysis are described through different SEDs and are expected to live in different haloes, therefore they require separate halo occupation distributions (HODs). The differential number counts are computed at each frequency band from the luminosity functions and the SEDs of each population.

   We note a difference with respect to the ERS case, for which the number counts at a reference frequency are the starting point. From an observational point of view, the millimetre counts of dusty galaxies are less determined than the radio counts at 5 GHz, while the infrared luminosity function is tightly constrained over a wide range of redshift by Herschel~\citep{berta2011building, bethermin2012hermes}, Spitzer~\citep{bethermin2010spitzer}, and Planck~\citep{negrello2013local}. Starting from the luminosity function allows for a straightforward computation of the number counts, without the need for a spectral extrapolation between a reference band and the observational bands.
   Therefore, for DSFG the reference frequency $\nu_*$ coincides with each measurement frequency $\nu$ (defined in Sect.~\ref{ssec:model.nccl}), unlike in the ERS case. 

    \subsubsection{Differential number counts from luminosity functions and SEDs}\label{sssec:model.DSFG.counts}
      For a population $t$ with bolometric infrared luminosity function $\Phi_t(\log_{10}L_{\rm IR}, z)$, the surface density of sources per unit flux and redshift at the observed frequency $\nu$ follows from Eqs.~8 and 12 of~\cite{cai2013hybrid}:
      \begin{equation}
        \frac{dn_t}{dz}\equiv\frac{d^3 N_t}{dS_\nu\, dz\, d\Omega}=
        \Phi_t\!\left(\log_{10}L_{\rm IR},z\right)\,
        \frac{d\log_{10}L_{\rm IR}}{dS_\nu}\frac{d^2 V}{dz\, d\Omega},
        \label{eq:dsfg_counts}
      \end{equation}
      where $d^2 V/dz\,d\Omega$ is the comoving volume element per unit solid angle, and the luminosity corresponding to a source of flux $S_\nu$ at redshift $z$ is 
      \begin{equation}
        L_{\rm IR}=
        \frac{4\pi D_L^2(z)\, S_\nu}{(1+z)\, \varphi_t\!\left(\nu(1+z)\right)}.
        \label{eq:dsfg_LIR}
      \end{equation}
      Here $L_{\rm IR}$ is the bolometric infrared luminosity, $D_L(z)$ is the luminosity distance, and $\varphi_t(\nu)$ is the SED template of population $t$ normalised to  $\int\varphi_t(\nu)\,d\nu=1$. We evaluate Eq.~\ref{eq:dsfg_counts} monochromatically at the nominal band centres and adopt best-fit \planck cosmological parameters to relate distances and redshifts, consistently with the halo model of Sect.~\ref{sssec:model.DSFG.clustering}.

      In the notation of Sect.~\ref{ssec:model.nccl}, for DSFG there is no spectral parameter dependence, but, contrary to the ERS case, the redshift dependence can not be neglected: the model assigns one deterministic SED to all sources of a given population at a given redshift $z$, so that the fluxes of a source at any two frequencies $\nu$ and $\nu_i$ entering the calculations are related by
      \begin{equation}
        \frac{S_{\nu_i}}{S_\nu}(z,t)=
        \frac{\varphi_t\!\left(\nu_i(1+z)\right)}{\varphi_t\!\left(\nu(1+z)\right)}.
        \label{eq:dsfg_sed_ratio}
      \end{equation}
      The average $\langle\cdot\rangle_{\boldsymbol{\theta}_t}$ appearing in equations \ref{eq:pdetect} and \ref{eq:xcls_BB} therefore involves no distribution of spectral parameters, and reduces to a deterministic factor, that remains inside the integral over redshift weighted by the counts.

    \subsubsection{Luminosity functions and SEDs of the two populations}\label{sssec:methods.DSFG.lfseds}
\begin{figure}
          \centering
          \includegraphics[width=\linewidth]{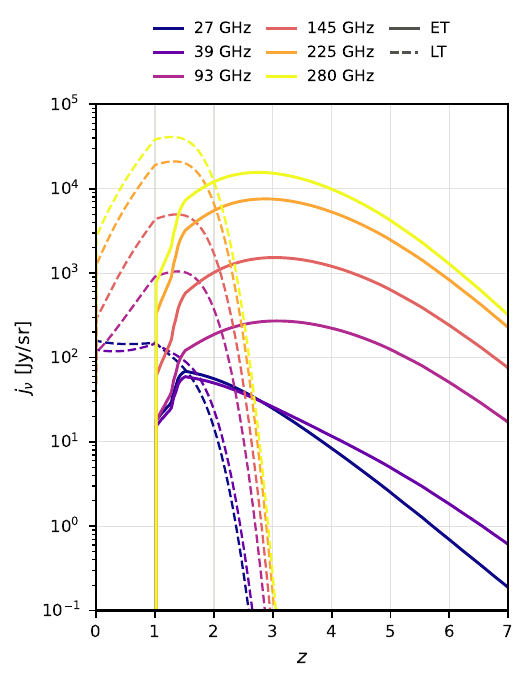}
          \caption{Emissivities $j_\nu(z)$ of dusty star-forming galaxies at the six SO frequency bands. Solid lines show early-type (ET) galaxies, dashed lines late-type (LT) galaxies. The ET emissivity vanishes below $z=1$, where proto-spheroids are not present in the model.}
          \label{fig:emiss_DSFG}
        \end{figure}

      Both populations are described by the bolometric infrared luminosity function of~\cite{cai2013hybrid}, as implemented by~\cite{zagatti2024halo}, to which we refer for details. For the LT galaxies, these are modified log-Gaussian functions following~\cite{saunders1990sixty}, whose characteristic density and luminosity evolve with redshift up to $z\simeq1$ and are then smoothly cut off. For the ET galaxies the luminosity function comes from the physical model of~\cite{granato2004physical}, and is truncated below $z\simeq1$ (see Fig.~\ref{fig:emiss_DSFG}).

      Each sub-population carries a single SED template: for the proto-spheroids, that of a strongly lensed $z\simeq2.3$ submillimetre galaxy \citep{swinbank2010intense, lapi2011herschel}; for the cold and warm late types, templates built from~\cite{dale2002infrared} and~\cite{smith2012herschel} at $\log_{10}(L_{\rm IR}/L_\odot)=9.75$ and $11.25$ respectively. At lower rest frequencies the templates also include the radio emission associated with star formation, namely synchrotron radiation from supernova remnants and free--free emission from H\textsc{ii} regions. This falling component, together with the rising dust one, produces a minimum in $S_\nu$ at a rest frequency of about 68~GHz. We stress that this emission is physically distinct from that of the ERS of Sect.~\ref{ssec:model.radio}, which originates in the jets and lobes powered by an active nucleus, so that adding the two models does not double count it. Nevertheless, assuming that the radio components of the DSFGs should be scaled by the same polarised fraction $\Pi$ as their thermal dust component is likely incorrect (certainly so in the case of the free-free contribution). As we will see, the impact of polarised DSFGs is significantly smaller than that of ERSs and thus, the choice to scale all the DSFG components equally is a reasonable approximation.

      Since each sub-population carries one template with no scatter in the SED, Eq.~\ref{eq:dsfg_sed_ratio} is fixed for a given $(t,z)$. The cross-frequency correlation coefficients predicted by the model are therefore upper limits: introducing a scatter in the SEDs would reduce the coherence between bands.

    \subsubsection{Power spectra and clustering}\label{sssec:model.DSFG.clustering}
      Unlike ERSs, DSFGs trace the large-scale structure coherently and contribute a potentially non-negligible clustering term in addition to the Poisson contribution. We therefore write the total DSFG power spectrum as
      \begin{equation}\label{eq:dsfg_clstot}
        C_\ell^{BB,\nu_1\nu_2}=C_\ell^{BB\,\rm PN,\nu_1\nu_2}+C_\ell^{BB\,\rm clus,\nu_1\nu_2},
      \end{equation}
      where the first (Poisson) term is Eq.~\ref{eq:xcls_BB} evaluated with DSFG counts, and the second is computed with the two-population halo model of~\cite{zagatti2024halo}.

      The quantity that carries the frequency dependence in the model is the emissivity (Fig.~\ref{fig:emiss_DSFG}), $j_t(\nu,z)\equiv dI_\nu/dz$, the contribution to the specific intensity of the unresolved background per unit redshift, in ${\rm Jy\,sr^{-1}}$ per unit $z$, so that $\int j_t(\nu,z)\,dz$ is the total intensity from population $t$. We evaluate it separately for each population from the counts of Eq.~\ref{eq:dsfg_counts}:
      \begin{equation}
        j_t(\nu,z)=\sum_{t\in P}\int dS_\nu\,\frac{dn_t}{dz}\,S_\nu\, f_\nu,  \label{eq:emissivity}
      \end{equation}
      with $t\in\{{\rm ET},{\rm LT}\}$. Here $f_\nu$ is the masking factor of Sect.~\ref{ssec:model.masking}. It is worth noticing that, in this case, the emissivity is dominated by faint sources, and is therefore essentially unaffected by the masking procedure. Most of the dependence on the masking strategy resides in the Poisson term. 

      Projecting the emissivities with the Limber approximation gives the angular power spectrum of the clustering part,
      \begin{equation}\label{eq:limber}
        C_\ell^{\rm clus, \nu_1\nu_2} =
        \sum_{t,t'}\int \frac{dz}{\chi^2}\, \frac{dz}{d\chi}\,
        j_t(\nu_1,z)\, j_{t'}(\nu_2,z)\,  P^{tt'}_{\rm gal}(k_\ell,z),
      \end{equation}
      with $k_\ell\equiv(\ell+1/2)/\chi$ in the Limber approximation~\citep{limber1953analysis, loverde2008extended}, and $P^{tt'}_{\rm gal}(k,z)$ the halo model galaxy power spectrum between sources of type $t$ and $t'$, including both the one-halo and two-halo contributions, built from the HODs of the two populations. We assigned different minimum hosting halo masses, $M_{\rm min}=10^{12}\,M_\odot h^{-1}$ for ET and $10^{11}\,M_\odot h^{-1}$ for LT, reflecting the fact that ET galaxies form in more massive structures. 

      Converting the intensity clustering into polarisation requires more than the $\langle\Pi^2\rangle/2$ factor alone used for the Poisson term, since we cannot assume random polarisation angles of physically associated sources. Following ~\cite{lagache2020impact} we write
      \begin{equation}
        C_\ell^{BB,\rm clus}=\frac{\langle\Pi^2\rangle}{2}
        \left[\rho_{\rm 1h}\, C_\ell^{\rm 1h}+
        \rho_{\rm 2h}\, C_\ell^{\rm 2h}\right],\qquad \rho \in [0,1],
        \label{eq:pol_clustering}
      \end{equation}
      where $\rho_{\rm 1h}$ and $\rho_{\rm 2h}$ quantify the degree of alignment of the polarisation angles within a halo and between distinct haloes, respectively. We set $\rho_{\rm 2h}=0$: galaxies in different haloes have no physical reason to share a preferred orientation, so the two-halo term does not contribute to the clustering polarisation power. We set $\rho_{\rm 1h}=1$, that is, we assume perfect alignment within a halo. This is an upper bound rather than a prediction. 

    \subsubsection{Summary of quantities computed by the model}\label{sssec:model.DSFG.summary}
      The model used for DSFGs takes as inputs
      \begin{equation}
        \left\{
        \Phi_t(\log_{10}L_{\rm IR},z),\,
        \varphi_t(\nu),\,
        f_\nu,\,
        \mathcal{P}(\Pi),\,
        {\rm HOD}_{\rm ET},\, {\rm HOD}_{\rm LT}
        \right\},
      \end{equation}
      where ${\rm HOD}_{\rm ET/LT}$ denotes the HOD model used to estimate the clustering of ET or LT galaxies. We use these ingredients to calculate the differential number counts of DSFGs (Eq.~\ref{eq:dsfg_counts}), and from that the two observables defined in  Sect.~\ref{ssec:model.nccl}, needed for this analysis:
      \begin{itemize}
        \item The angular number density of DSFGs whose flux at the masking frequency exceeds $S_{\rm lim}$. In the notation used in this section, this reads
        \begin{equation}\nonumber
          N^t_{\nu_m}(>S_{\rm lim})
          =\iint dS_\nu\,dz\,\frac{dn_t}{dz}\,\Theta\left[S_\nu\,\frac{\varphi_t(\nu_m(1+z))}{\varphi_t(\nu(1+z))} > S_{\rm lim}\right],
        \end{equation}
        with the total obtained as $N_{\nu_m}=\sum_t N^t_{\nu_m}$.
        \item The $BB$ cross-spectra, now with both a Poisson and a clustering contribution, as in Eq.~\ref{eq:limber}. Both terms are built from the same counts, but enter with a different power of the flux. The Poisson term weights each source by the product of its fluxes in the two bands, 
        \begin{align}
          C_\ell^{BB\,{\rm PN},\nu_1\nu_2}
          =&\;
          \frac{\langle\Pi^2\rangle}{2}
          \sum_t\iint dS_\nu\,dz\,\frac{dn_t}{dz}f_{\nu_1}S_{\nu_1}\,f_{\nu_2}S_{\nu_2},\label{eq:dsfg_shot_final}
        \end{align}
        with $S_{\nu_2}$ and $S_{\nu_1}$ obtained from $S_\nu$ from Eq.~\ref{eq:dsfg_sed_ratio}. The clustering contribution is given by Eq. \ref{eq:dsfg_clstot} scaled by $\langle \Pi^2\rangle/2$, with $\rho_{\rm 1h}=1$ and $\rho_{\rm 2h}=0$.
      \end{itemize}

      We use the polarisation fraction of ~\cite{dezotti2018exploring}, used also by~\cite{lagache2020impact}, $\langle\Pi_{\rm IR}\rangle=1.4\%$, independent of frequency and flux, and approximate $\langle\Pi^2\rangle$ by $\langle\Pi\rangle^2$. As for the ERS, the results are converted from ${\rm Jy^2\,sr^{-1}}$ to $\mu{\rm K^2_{CMB}}$ with the monochromatic conversion factors evaluated at band centres.

\section{Analysis framework}\label{sec:analysis}
  We propagate the impact of point source contamination by generating synthetic cross-frequency power spectrum data containing CMB and Galactic foreground contributions as well as the residual PS contribution calculated as described in Section \ref{sec:model}. We then derive constraints on cosmological and foreground parameters through a power spectrum-based component separation and likelihood framework. The impact of point sources is then reflected in the bias of the inferred parameters (caused by the residual PS contamination), and in their statistical uncertainties (caused largely by the loss of sky area due to masking).

  \subsection{Synthetic data}\label{ssec:analysis.synth}
    We build synthetic data vectors of cross-frequency polarisation power spectra. The models used to simulate the CMB and Galactic components are as follows:

    \begin{enumerate}
      \item {\bf The CMB} signal is generated with \texttt{CAMB}~\citep{lewis2000efficient} for a $\Lambda$CDM model with the best-fit cosmological parameters from \planck. We fix the tensor-to-scalar ratio to $r=0$ (no primordial $B$ modes), and rescale the $B$-mode power spectrum due to gravitational lensing by an amplitude $A_\mathrm{lens}=0.3$. This represents the level of delensing ($70\%$) that could be achievable by an experiment such as SO \citep{hertig2024simons}.
      \item {\bf The dust and synchrotron} components are modelled as having a power-law power spectrum of the form
      \begin{equation}
        {\cal D}_\ell^x\equiv\frac{\ell(\ell+1)}{2\pi}C_\ell^x=A_x\left(\frac{\ell}{\ell_0}\right)^{\alpha_x},
      \end{equation}
      with $\ell_0=80$. This is the power spectrum for component $x$ (with $x\equiv d$ for dust and $x\equiv s$ for synchrotron) at a pivot frequency $\nu_0^x$. We use $\nu_0^s=23\,{\rm GHz}$ and $\nu_0^d=353\,{\rm GHz}$. These power spectra are scaled to other frequencies assuming a modified black-body spectrum for dust and a power-law spectrum for synchrotron:
      \begin{equation}
        f_d(\nu)=\left(\frac{\nu}{\nu_0^d}\right)^{\beta_d}\frac{B_\nu(T_d)}{B_{\nu_0}(T_d)},
        \hspace{12pt}
        f_s(\nu)=\left(\frac{\nu}{\nu_0^s}\right)^{\beta_s}.
      \end{equation}
      Note that both of these spectra are defined in Rayleigh-Jeans temperature units. We use the following values for the foreground parameters defining these contributions:
      \begin{eqnarray}\nonumber
        &\beta_s=-3,\hspace{6pt}\beta_d=1.54,\hspace{6pt}\alpha_s=-0.8,\hspace{6pt}\alpha_d=-0.4,\\
        &A_s=1\,\mu K^2_{\rm CMB},\hspace{6pt}A_d=22\,\mu K^2_{\rm CMB},\hspace{6pt}T_d=20\,\mathrm{K}.
      \end{eqnarray}
      These values are representative of the amplitude, scale dependence, and frequency spectra of both components in the {\tt d0} and {\tt s0} models implemented in {\tt PySM} \citep{thorne2017python} within the angular footprint of an SO-like (i.e. Chile-based) experiment.
    \end{enumerate}
    We note that the true Galactic sky is likely more complex than this, with each foreground component not exactly described by a power law $\mathcal{D_\ell}$ model and more complex spectral dependence, with foreground SEDs varying across the Galaxy, both across the sky and along any given line-of-sight \citep[e.g.][]{krachmalnicoff2018s, akrami2020planck, martire2022characterization, weiland2022polarized}. However, considering such a simple model allows us to explore an optimistic scenario recovering the tightest possible constraints on $r$ as well as all other model parameters, making the results more sensitive to the impact of residual point source contamination. 
    The impact and performance of more complex foreground parametrisations on $B$-mode constraints have been studied in the literature \citep[e.g.][]{azzoni2021minimal,vacher2022moment,Wolz:2023lzb,liu2025simons}.

  \subsection{Error model}\label{ssec:analysis.error}
    \begin{figure}
      \centering
      \includegraphics[width=\linewidth]{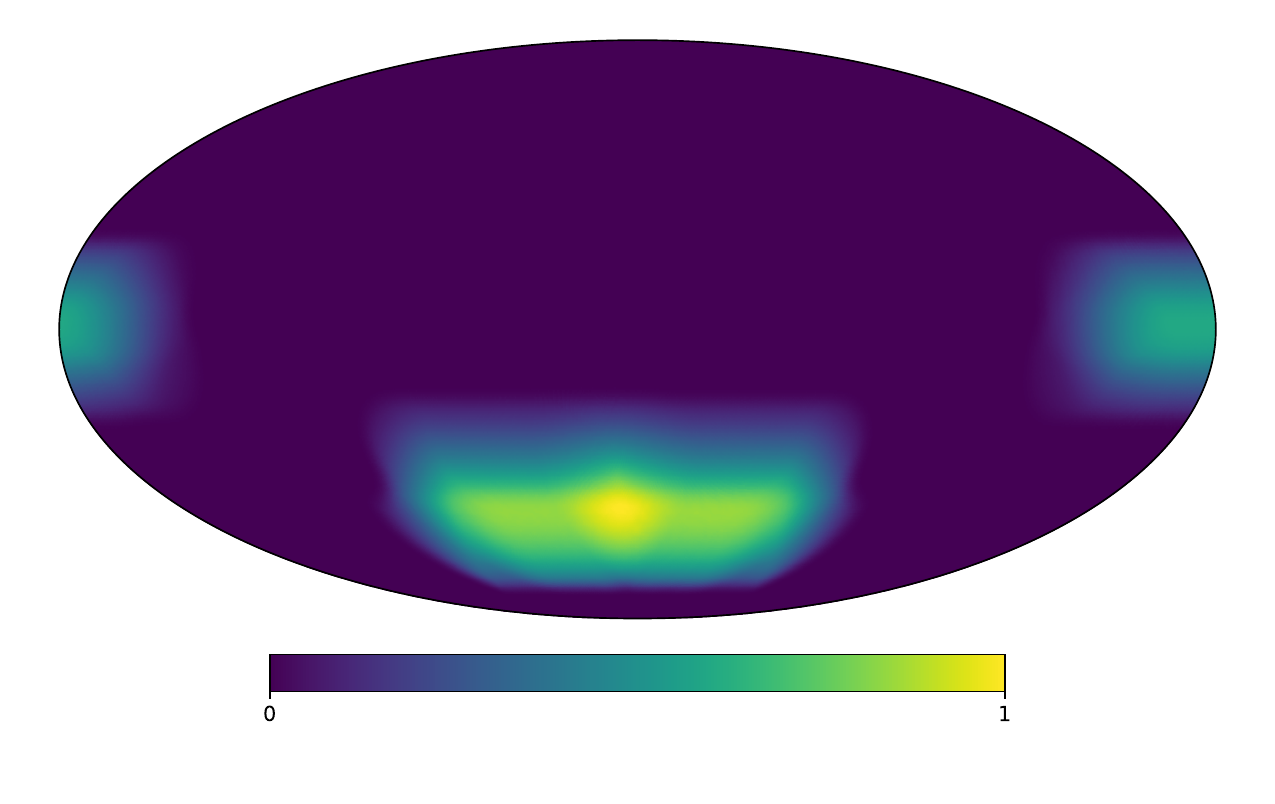}
      \caption{Footprint of a SO-like Chile-based experiment used in this analysis, shown in equatorial coordinates.}
      \label{fig:mask}
    \end{figure}
    \begin{table}
      \centering
      \begin{tabular}{cccccc}
      \hline
      Freq. [GHz] & \shortstack{$\mu \mathrm{K}$-arcmin \\ baseline} & \shortstack{$\mu \mathrm{K}$-arcmin \\ extended} &
      $\ell_\mathrm{knee}$ & $\alpha_\mathrm{knee}$ & FWHM [$'$] \\
      \hline
      27  & 49.5 & 12.0 & 30  & 2.4 & 91 \\
      39  & 29.7 & 7.7  & 30  & 2.4 & 63 \\
      93  & 3.7  & 1.2  & 50  & 2.5 & 30 \\
      145 & 4.7  & 1.4  & 50  & 3.0 & 17 \\
      225 & 8.9  & 3.9  & 70  & 3.0 & 11 \\
      280 & 22.6 & 9.6  & 100 & 3.0 & 9  \\
      \hline
      \end{tabular}
      \caption{Frequency bands and instrumental specifications assumed for this analysis. We list the white-noise amplitudes for the baseline and extended noise configurations, the parameters describing correlated ``$1/\ell$'' noise, and the FWHM resolution of each channel.}\label{tab:noise_values}
    \end{table}
    The synthetic data described in the previous section are generated without any measurement noise component (neither due to cosmic/sample variance nor instrumental noise). This ensures that any deviation in the inferred model parameters from the input values used to generate the synthetic data represents a genuine bias caused by the presence of point source contamination. Nevertheless, in order to obtain representative statistical uncertainties on the model parameters, against which any such biases may be compared, we calculate the covariance matrix of the synthetic data accounting for all sources of measurement uncertainty, including the impact of point source masking.

    The power spectrum covariance is calculated assuming all fields involved are Gaussian, as is standard in most CMB power spectrum analyses \citep{hamimeche2008likelihood,Planck:2015bpv,atkins2025atacama}\footnote{Note that foreground non-Gaussianity does not lead to significant parameter biases if ignored in the covariance matrix \cite{abril2024impact}.}. To account for the impact of mode-coupling due to the presence of a finite sky mask we use the improved Narrow Kernel Approximation \citep{garcia2019disconnected,nicola2021cosmic} as implemented in {\tt NaMaster}\footnote{\url{https://github.com/LSSTDESC/NaMaster}} \citep{alonso2019unified}.
    \begin{figure}
      \centering
      \includegraphics[width=\linewidth]{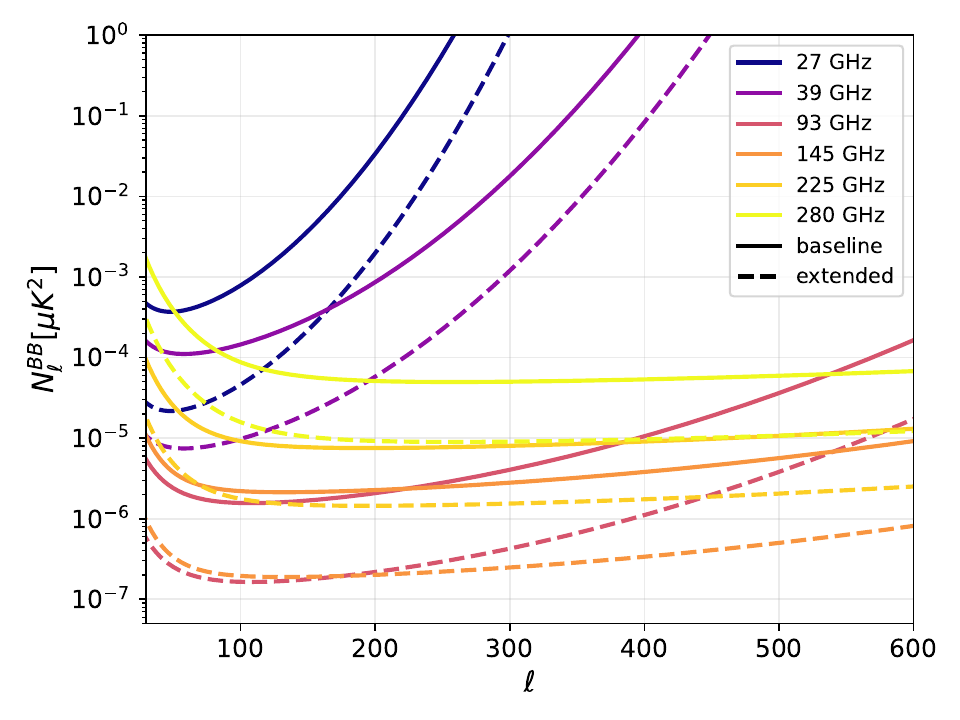}
      \caption{Noise spectra for the "baseline" and "extended" configurations studied. The $1/\ell$ noise component is visible at the largest scales, while the contribution due to deconvolving the instrumental beam dominates at progressively smaller scales going to higher frequency.}
      \label{fig:noise_spectra}
    \end{figure}
    
    As a mask representative of the sky region targeted by current ground-based experiments, we use the publicly-available hits map of the SO Small Aperture Telescopes \citep[SATs,][]{ade2019simons}, shown in Fig. \ref{fig:mask}. To account for the impact of PS masking in the covariance matrix, we re-scale the covariance matrix obtained in the absence of any PS mask by sky area remaining after masking:
    \begin{equation}
      {\rm Cov}_{\rm with\,PS}=\frac{f_{\rm sky,no\,PS}}{f_{\rm sky,with\,PS}}\,{\rm Cov}_{\rm no\,PS}.
    \end{equation}
    Here $f_{\rm sky,no\,PS}$ is the effective area of the original mask ($f_{\rm sky,no\,PS}=0.097$ in the case of the mask shown in Fig. \ref{fig:mask}). The available sky fraction after point source masking is given by
    \begin{equation}
      f_{\rm sky,with\,PS}=f_{\rm sky,no\,PS}\exp\left[-N_{\nu_m}(>S_{\rm lim})\Omega_{\rm mask}\right],
    \end{equation}
    where $N_{\nu_m}(>S_{\rm lim})$ is the angular number density of masked point sources (see Eq. \ref{eq:ncounts_per_pop}), and $\Omega_{\rm mask}=\pi R_{\rm mask}^2$ is the solid angle associated with the mask of a single point source (with $R_{\rm mask}$ given in Eq. \ref{eq:r_mask}).

    We include the contribution from instrumental noise in the covariance calculation. For this, we use the noise model presented in \cite{ade2019simons}, with
    \begin{equation}
      N_\ell = N_\mathrm{white} \left( 1 + \left(\frac{\ell}{\ell_\mathrm{knee}}\right)^{\alpha_\mathrm{knee}}\right).
    \end{equation}
    Here $N_{\rm white}$ is the white noise amplitude, and $\ell_{\rm knee}$ and $\alpha_{\rm knee}$ characterise the non-white (``$1/\ell$'') component, largely associated with the atmosphere. These three parameters vary across frequency channels. We will consider an SO-like experiment, with 6 frequency bands listed in Table \ref{tab:noise_values}, together with the values of $N_{\rm white}$, $\ell_{\rm knee}$, and $\alpha_{\rm knee}$ considered in this work, as well as the beam FWHM apertures. To model the $1/\ell$ noise component we use the values of $\ell_{\rm knee}$ and $\alpha_{\rm knee}$ corresponding to the ``pessimistic'' model of \cite{ade2019simons}. We investigate two white noise scenarios, the ``baseline'' noise level of \cite{ade2019simons}, corresponding to a 5-year experiment, and the ``extended'' configuration of \cite{abril2026simons}, with up to $\sim3$ times higher sensitivity. The resulting noise spectra, in both configurations, are shown in Fig. \ref{fig:noise_spectra}, including the impact of deconvolving the instrumental beam.

    When considering ERSs we will assume they are masked in the lowest frequency channel (i.e. $\nu_m=27\,{\rm GHz}$), where the beam FWHM is $91'$, as this is the channel with more and brighter sources. Masking radio sources may thus lead to a significant loss of area. 
    A different choice could be made for the DSFGs, however we do not explore this as we find that no source masking is required for DSFGs in our analysis.

  \subsection{Component separation and likelihood}\label{ssec:analysis.likelihood}
    We use a power spectrum-based Bayesian framework to separate the CMB and foreground components and derive constraints on the tensor-to-scalar ratio and other model parameters. The methodology is described in detail in e.g. \cite{Wolz:2023lzb} (see Pipeline A), and we only outline its main elements here. Cross-frequency $B$-mode power spectra are modelled as a linear combination of contributions from different components:
    \begin{eqnarray}
      C_\ell^{\nu_1\nu_2}=\sum_xC_\ell^x\,f_x(\nu_1)f_x(\nu_2).
    \end{eqnarray}

    The components considered include the CMB $B$-mode signal, Galactic thermal dust and synchrotron. The Galactic foregrounds are modelled as described in Section \ref{ssec:analysis.synth} in terms of both scale and frequency dependence. The CMB contribution is described in terms of $r$ and $A_{\rm lens}$, which multiply tensor-only and lensing-only templates respectively, calculated fixing all other $\Lambda$CDM cosmological parameters to the \planck best-fit values. The free parameters of the fiducial model are then
    \begin{eqnarray}
      \{r,\,A_{\rm lens},\,A_{d},\,A_s,\,\beta_d,\,\beta_s,\,\alpha_d,\,\alpha_s\}.
    \end{eqnarray}
    In addition to this minimal model, we will also consider the possibility of absorbing any residual point source contamination in terms of an additional component, labelled $p$, with an effective SED modelled as a power law in Rayleigh-Jeans temperature units $f_p(\nu)=(\nu/\nu^p_0)^{\beta_p}$ and with a white angular power spectrum characterised by a free amplitude $A_p$ (i.e. $C_\ell^p=A_p\,\forall\,\ell$). Finally, we have also tested the interplay between residual extragalactic PS contamination and additional Galactic foreground complexity, by allowing for frequency decorrelation as modelled in \cite{2018PhRvL.121v1301B} and checking that such a term is not detected.

    We approximate the power spectrum-level likelihood as a multivariate Gaussian distribution, which has been shown to be a good approximation on the scales explored here \citep{Wolz:2023lzb}. We adopt broad, largely uninformative priors on all model parameters, although we enforce the positivity of all foreground amplitude parameters ($A_d$, $A_s$, $A_p$). The model parameters are then recovered via Metropolis-Hastings sampling using Monte Carlo Markov Chains (MCMC). This multi-frequency power spectrum approach has been extensively used in CMB analysis, e.g.~\cite{planck2020planck, ade2021improved,louis2025atacama}, as well as in forecast papers, e.g.~\cite{Wolz:2023lzb, azzoni2021minimal}. We use the specific implementation of this method available in \texttt{BBPower}\footnote{\url{https://github.com/simonsobs/BBPower}}.

    The synthetic data vector described in Section \ref{ssec:analysis.synth} consists of all auto- and cross-correlations between the 6 frequency channels listed in Table \ref{tab:noise_values}, sampled in multipole bandpowers of width $\Delta\ell=10$ in the range $32<\ell\leq600$.

\section{Results}\label{sec:res}
  We first present results regarding the impact of residual contamination from polarised ERSs, as the impact of DSFG contamination on $B$ modes is significantly smaller. We present results for DSFGs in Section \ref{ssec:res.dsfg}.

  \subsection{ERS forecasts}\label{ssec:res.ersfore}
    \subsubsection{Number counts}\label{sssec:res.ersfore.nc}
\begin{figure}
        \centering
        \includegraphics[width=\linewidth]{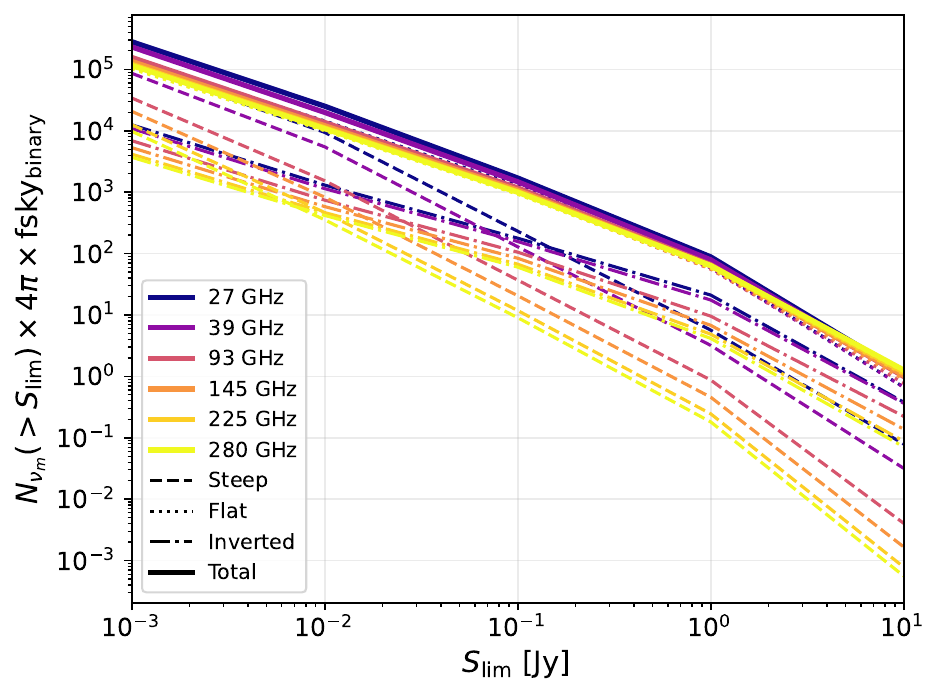}
        \caption{ ERS number counts in the sky patch shown in Fig. \ref{fig:mask} for different masking frequencies $\nu_m$ and as a function of $S_\mathrm{lim}$, for the different ERS populations. The contribution from flat sources dominates the total counts, and is largely hidden by the solid lines showing the sum of all populations.}
        \label{fig:num_counts_ERS_as_function_of_Slim_and_nu_mask}
      \end{figure}

      Using Eq.~\ref{eq:ncounts_per_pop}, we compute the ERS angular number density above the flux cut $S_\mathrm{lim}$, $N^t _\nu (> S_{\mathrm{lim}})$, multiplied by the SO SAT sky area, at the different SAT frequencies $\nu$. The result is shown in Fig.~\ref{fig:num_counts_ERS_as_function_of_Slim_and_nu_mask}.
      As expected, the number counts are largely dominated by the flat-spectrum sources. Moreover, for a fixed $S_\mathrm{lim}$ the number of sources increases towards lower frequency, motivating the choice of $\nu_m=27 \, \mathrm{GHz}$. We note that also the inverted spectrum sources follow this behaviour, since their number count is dominated by sources whose spectra peak before the SO lowest frequency of 27~GHz.
        
      In our analysis, these number counts are relevant for determining the sources to mask. We choose to identify the ERSs at the instrument lower frequency of 27~GHz (common both to the SATs and the LAT instruments). In principle, ERSs could be identified at even lower frequencies leveraging external data. However, we do not consider this necessary, as extending the source identification to lower frequency would increase the uncertainty associated with extrapolating the source frequency scaling over broad frequency ranges. Moreover, the choice of $\nu_{m}$ is not determined solely by maximising the number counts, but also by choosing at what frequency to prioritize the ERS masking, as the brightest sources at a given frequency are possibly not the same at other frequencies. We hence choose to prioritize the masking at the lowest frequency band, as that is the one where we have higher ERS signal (but not necessary higher point source signal-to-noise ratio, as we show later). Finally, the highest-frequency bands are not well suited for identifying the ERS to be masked, since thermal dust emission becomes increasingly important at these frequencies and could dominate the observed flux.

    \subsubsection{Angular power spectra}\label{sssec:res.ersfore.cls}
\begin{figure}
        \centering
        \includegraphics[width=\linewidth]{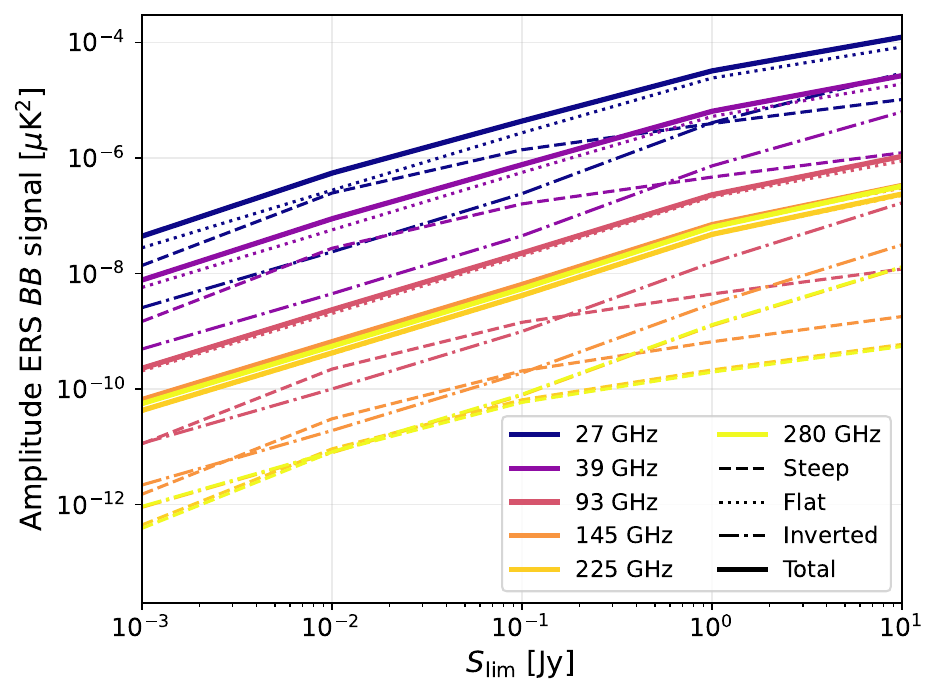}
        \caption{ERS power spectrum amplitude for different measurement frequencies $\nu$, with masking frequency $\nu_m = 27\,\mathrm{GHz}$ and $f_\mathrm{FWHM}=3$ as a function of $S_\mathrm{lim}$, shown for the different ERS populations and their sum.}
        \label{fig:ampl_PS_as_function_of_Slim_and_nu_mask}
      \end{figure}

      Having chosen the masking frequency $\nu_m$, we can then compute the amplitude of the residual ERS power spectrum after masking with different flux cuts $S_\mathrm{lim}$. Fig.~\ref{fig:ampl_PS_as_function_of_Slim_and_nu_mask} shows the amplitude of the ERS signal when masking with $\nu_m=27 \, \mathrm{GHz}$, and $f_\mathrm{FWHM}=3$, which means each masked PS is removed from the map almost completely (less than 0.2\% of the flux remains). This conservative approach may be accurate if we are only interested in suppressing the residual PS amplitudes, but may not be the best trade-off when also accounting for the noise and the fraction of sky lost due to masking (see Section~\ref{sssec:res.ers.maskrad}).

    \subsection{Residual ERS contamination and component separation}\label{ssec:res.ers}
      \subsubsection{Per-band masking}\label{sssec:res.ers.perband}
        We start by considering the flux limits and number counts studied in \cite{lagache2020impact}. This allows us to get a sense of what flux limits may be reasonably expected in future experiments, as well as a rough estimate of the parameter biases caused by different levels of PS contamination. Note that, although \cite{lagache2020impact} assumed that each frequency channel would be masked according to the sources detected in it by the instrument being analysed, we will show results for a uniform mask applied across frequencies, built from the sources detected in a single channel for its corresponding flux limit. The flux limits estimated by \cite{lagache2020impact} for the SO SAT and Large Aperture Telescope (LAT) bands are shown in Table~\ref{tab:num_counts_at_the_different_SAT_bands}, together with the associated expected number counts. In particular, at 27~GHz they find $S_\mathrm{lim}^\mathrm{SAT \, 27 \, GHz} = 1.983 \, \mathrm{Jy}$, growing at $39\,{\rm GHz}$ due to the decaying nature of radio spectra, but decreasing afterwards due to the better sensitivity and smaller beam sizes.
        \begin{table}
          \centering
          \begin{tabular}{c|cc|cc}
            \hline
            & \multicolumn{2}{c|}{SAT} & \multicolumn{2}{c}{LAT} \\
            \cline{2-5}
            $\nu$ [GHz]
            & $S_{\rm lim}$ [Jy]
            & $N^t_\nu(>S_{\mathrm{lim}})$
            & $S_{\rm lim}$ [Jy]
            & $N^t_\nu(>S_{\mathrm{lim}})$ \\
            \hline
            27  & \textbf{1.983}  & \textbf{27}  & \textbf{0.01502} & $\mathbf{1.565\times10^{4}}$ \\
            39  & 4.177  & 6   & 0.00994 & $2.004\times10^{4}$ \\
            93  & 3.385  & 8   & \textbf{0.00296} & $\mathbf{5.028\times10^{4}}$ \\
            145 & 1.593  & 31  & 0.00261 & $4.966\times10^{4}$ \\
            225 & 0.7558 & 94  & 0.00568 & $2.000\times10^{4}$ \\
            280 & \textbf{0.4351} & \textbf{191} & 0.01011 & $1.067\times10^{4}$ \\
            \hline
          \end{tabular}
          \caption{For the baseline configuration, number of ERS per steradian above $S_{\rm lim}$. Where the $S_{\rm lim}$ are as computed by \protect\cite{lagache2020impact} for a SAT-like and a LAT-like instrumental setup. Written in bold are: the 27~GHz case, as the scenario at which we would most naturally mask the sources, the 280~GHz band for the SAT and the 93~GHz band for the LAT, as the bands at which the largest number of sources are nominally detected (even if realistically one would not rely on these high frequency bands, as there would be a significant contribution from infrared dust emission not taken into account here).}\label{tab:num_counts_at_the_different_SAT_bands}
        \end{table}

\begin{figure}
          \centering
          \includegraphics[width=\linewidth]{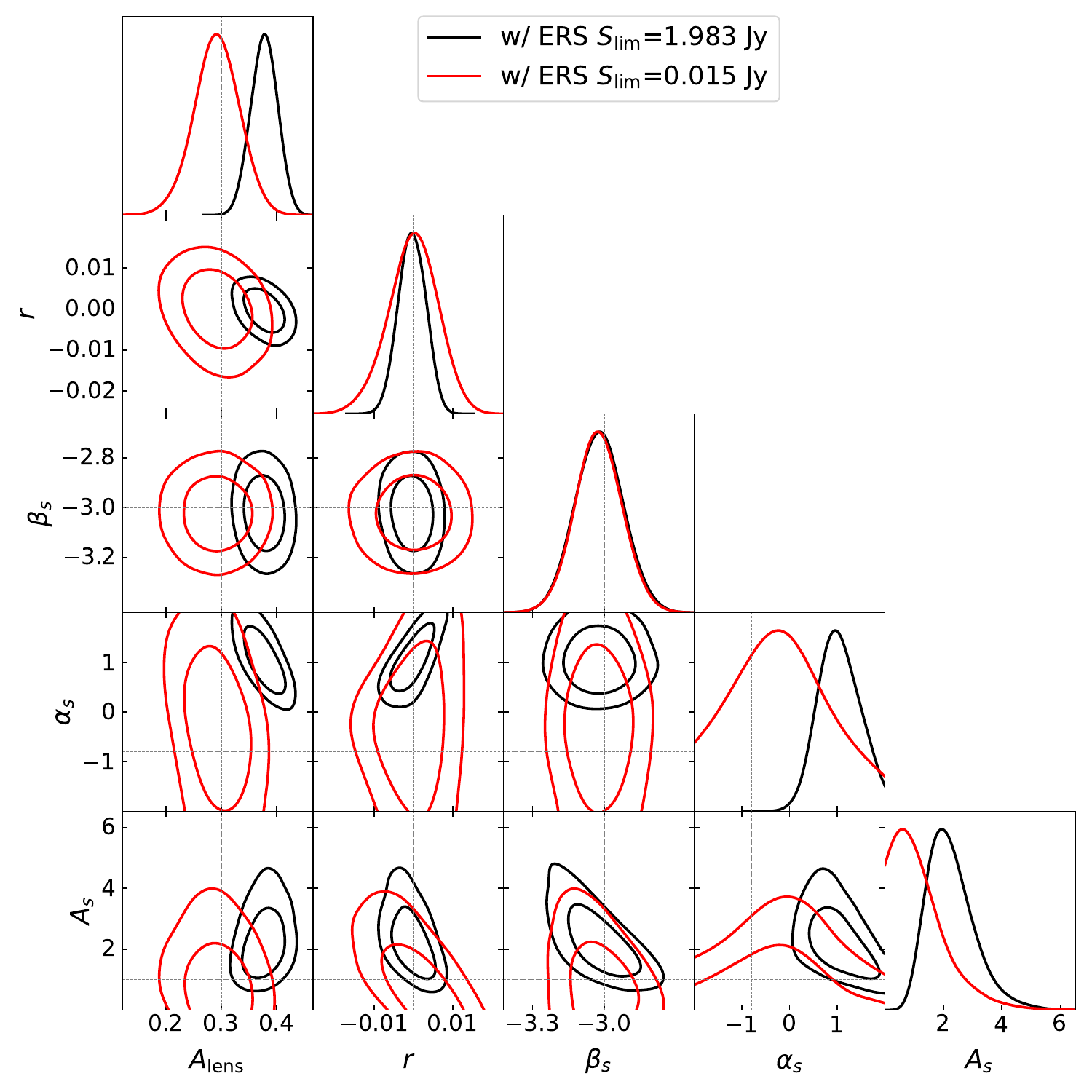}
          \caption{For the baseline configuration, we show the marginalised posteriors for the CMB and synchrotron parameters, recovered with the SAT and LAT baseline $S_\mathrm{lim}$ at $\nu_m=\mathrm{27 \, GHz}$. The Galactic dust parameters are unbiased in both cases and are not shown here. Note that when masking with the less stringent SAT $S_\mathrm{lim}=1.983\,\mathrm{Jy}$ we find no significant difference when varying $f_\mathrm{FWHM}$ from 1 to higher values, while for the LAT case $S_\mathrm{lim}=0.015\,\mathrm{Jy}$ we show the result for $f_\mathrm{FWHM}=1$. Increasing the masking radius further would lead to a larger degradation in the recovered parameter uncertainties.}
          \label{fig:Slim_SAT_baseline}
        \end{figure}

        Adopting the flux limit found for the SATs at 27 GHz, and propagating the impact of the residual contamination to the model parameters, we find the results shown by the black contours of Fig.~\ref{fig:Slim_SAT_baseline}. We find that the synchrotron parameters, particularly the tilt $\alpha_s$ and amplitude $A_s$, as well as the CMB lensing amplitude $A_{\rm lens}$, are significantly biased. This is not surprising, as synchrotron is the Galactic foreground that most closely matches the SEDs of radio sources, while the scale dependence of the lensing $B$-mode component is largely flat on the scales ($\ell<600$) studied here, similar to Poisson sources. Fortunately, the tensor-to-scalar ratio $r$ remains unbiased. Even at the more stringent flux limit found at 280 GHz (even though masking at this high frequency is not realistic given the associated emission from dust in the same sources, disregarded here), we find that these parameter biases do not disappear.
        This analysis is performed with $f_{\rm FWHM}=1$, but because of the high $S_\mathrm{lim} ^\mathrm{27 \, GHz}$ the result does not change with $f_{\rm FWHM}=3$.
        
         We may also consider the case where the positions of sources to be masked come not from the same instrument being analysed, but from an external dataset, more sensitive to point source emission (e.g. because of better resolution). In particular, let us use the detection limit for the SO LAT at 27 GHz, more than 100 times smaller than the SAT flux limit at the same frequency. At this limit we are able to detect and mask a much larger number of sources, reducing the parameter bias but also the size of the sky area available for the analysis. We find that the parameter biases reported earlier disappear completely, albeit at the cost of a significant increase in the posterior parameter uncertainties (by a factor 1.8 in the case of $r$). While the discussion in this section has focused on the baseline configuration similar results hold with the higher sensitivity of the extended configuration.

        For the remainder of this paper we will therefore explore a range of flux limits, spanning the fluxes available to SAT-like and LAT-like experiments, to study whether a suitable trade-off between these two effects (reduced bias and increased variance) can be achieved. Specifically, we will study three cases: $S_{\rm lim}=0.01 \, \mathrm{Jy}$, $0.1 \, \mathrm{Jy}$ and $1 \, \mathrm{Jy}$.

    \subsubsection{Choice of the masking radius for each ERS}\label{sssec:res.ers.maskrad}
      \begin{figure}
        \begin{minipage}[b]{0.48\textwidth}
          \centering
          \includegraphics[width=\linewidth]{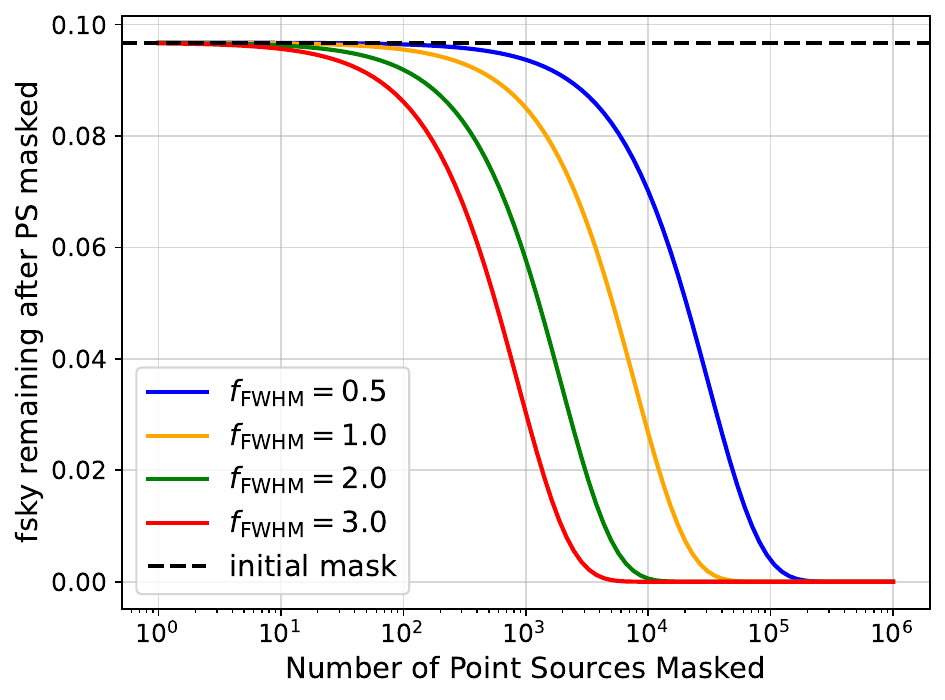}
        \end{minipage}
        \begin{minipage}[b]{0.48\textwidth}
          \centering
          \includegraphics[width=\linewidth]{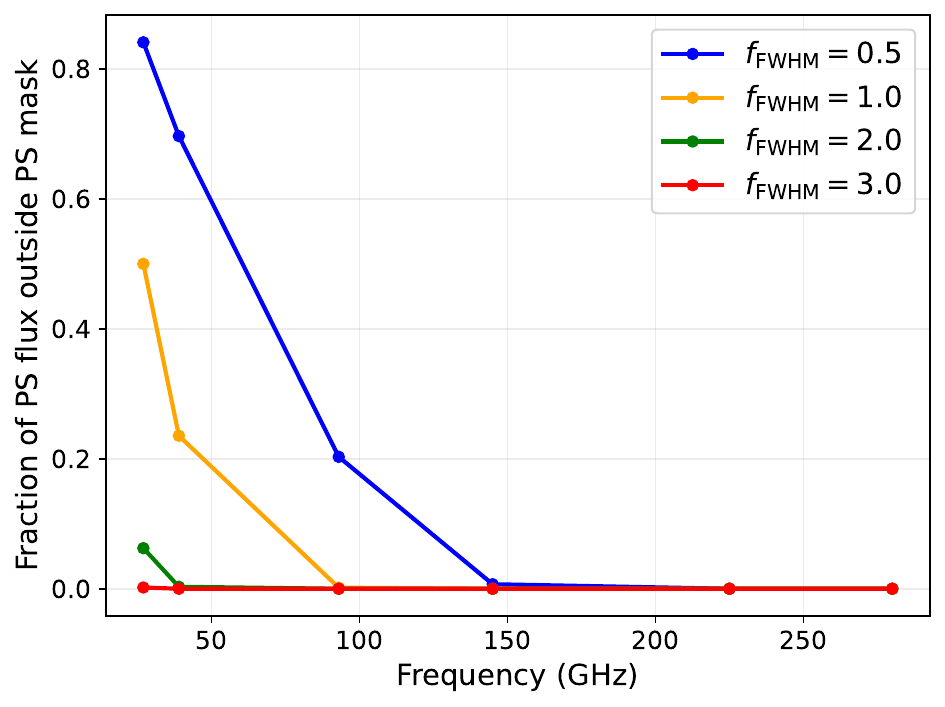}
        \end{minipage}
        \caption{\textit{Top figure}: surviving sky fraction after point source masking as a function of the number of masked point sources, for 3 different choices of radius for the circular mask on each source. The radius is taken as half of the FWHM of the beam at 27~GHz times a factor $f_\mathrm{FWHM}$. Masking each source with a circular hole of radius given by 1 FWHM of the beam at 27~GHz ($91'$) ($f_\mathrm{FWHM}=2$) would mean having a significant loss in sky fraction (from 0.097 to 0.06) when masking 1000 sources. \textit{Bottom figure}: residual point source flux after masking for the same 3 choices of $f_{\rm FWHM}$, at the different analysis frequencies. With $f_{\rm FWHM}=1$ at 27~GHz ($91'$), half of the PS flux is left unmasked at that frequency, a fraction that drops steeply towards higher frequencies.}\label{fig:f_fwhm_choices}
      \end{figure}
      The choice of masking strategy depends not only on the flux limit used to detect and mask sources, but also on the aperture size used for masking. In our analysis we assume that sources are masked at the lowest frequency $\nu_m = 27 \, \mathrm{GHz}$, where $\theta_{\mathrm{FWHM}}(\nu_m)=91'$. The choice of masking fraction $f_{\rm FWHM}$ (which determines the masking radius -- Eq.~\ref{eq:r_mask}) in this case is not trivial: large values of $f_{\rm FWHM}$ ensure that sources are fully masked at the masking frequency $\nu_m$, and certainly so at all other (higher) frequencies, where the beam is smaller, although this is at the cost of an increased loss of usable area. Choosing a smaller $f_{\rm FWHM}$ thus reduces this loss, increasing the residual contamination at $\nu_m$ but perhaps not so much at the higher frequencies (e.g. 90-150 GHz), where the beam is smaller and the overall sensitivity is typically higher. We explore different choices of $f_{\rm FWHM}$:
      \begin{enumerate}
        \item $R_\mathrm{mask}(f_\mathrm{FWHM}=0.5) = 22.75'$,
        \item $R_\mathrm{mask}(f_\mathrm{FWHM}=1) = 45.5'$,
        \item $R_\mathrm{mask}(f_\mathrm{FWHM}=2) = 91'$,
        \item $R_\mathrm{mask}(f_\mathrm{FWHM}=3) = 136.5'$.
      \end{enumerate}
      We study the trade-off between area loss and residual PS contamination of these choices in Fig.~\ref{fig:f_fwhm_choices}. We find that $f_{\rm FWHM}=1$ (corresponding to a circular mask the size of the beam FWHM at $\nu_m=27\,{\rm GHz}$) achieves a somewhat reasonable balance: this choice allows us to mask up to 1000 sources across the sky without an excessive loss of sky area and, at the same time, although half of the masked PS flux remains at $\nu_m$, it vanishes almost completely at the more sensitive 90 and 150 GHz bands. Other choices lead to either a significantly larger sky loss or a non-negligible contamination at intermediate frequencies.
        
\begin{figure}
        \centering
        \includegraphics[width=\linewidth]{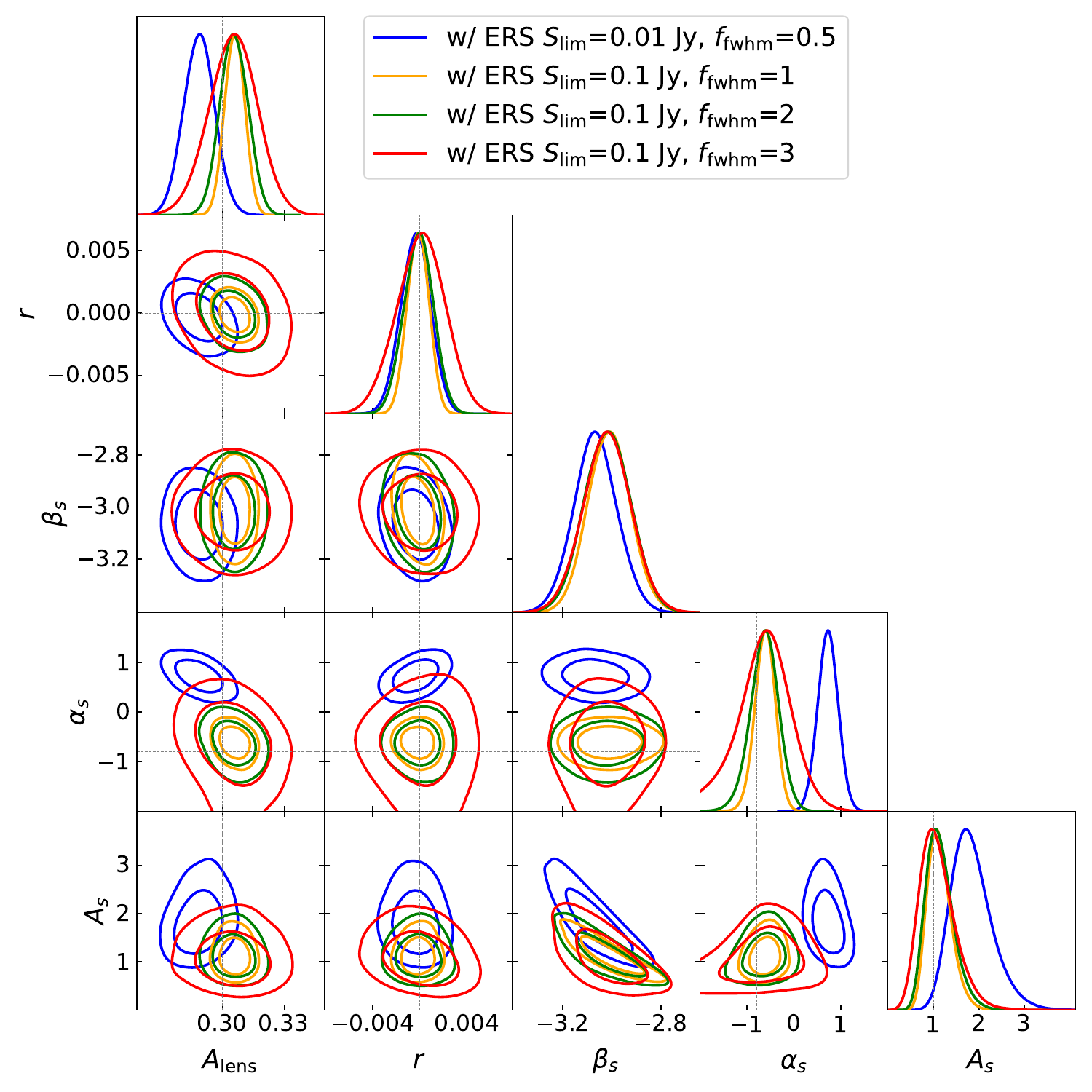}
        \caption{Parameter constraints found for the extended noise configuration for different values of the masking radius $f_{\rm FWHM}$.}
        \label{fig:choice_fFWHM_triangle}
      \end{figure}

      To quantify the impact of this choice more rigorously, we study the parameter constraints obtained with each of the options above. We assume the extended configuration in this case. The result is shown in Fig.~\ref{fig:choice_fFWHM_triangle}. For simplicity we adopt a flux cut of $S_{\rm lim}=0.1\,{\rm Jy}$ in all cases but $f_{\rm FWHM}=0.5$, where a more stringent cut of $S_{\rm lim}=0.01$ may be adopted with only a mild loss of area. We find that the fiducial case of $f_{\rm FWHM}=1$ achieves the best balance of all explored: it achieves the smallest statistical uncertainties while incurring virtually no bias on the model parameters. The mild $\sim1\sigma$ bias on $A_{\rm lens}$ (which we will return to in Section~\ref{sssec:res.ers.marg}) is still apparent in the more conservative cases $f_{\rm FWHM}=2$ and 3, showing that it is driven by the residual contribution from unmasked sources, and not by the residual flux of the masked ones. Interestingly, even with the more stringent flux cut, the case $f_{\rm FWHM}=0.5$ is significantly more biased, showing that, in this case, the dominant effect is the residual flux of the masked sources which, as we showed in Fig. \ref{fig:f_fwhm_choices}, extends to the high-sensitivity $90\,{\rm GHz}$ channel.

    \subsubsection{Masking: bias-uncertainty trade-off}\label{sssec:res.ers.tradeoff}
\begin{figure}
        \centering
        \includegraphics[width=\linewidth]{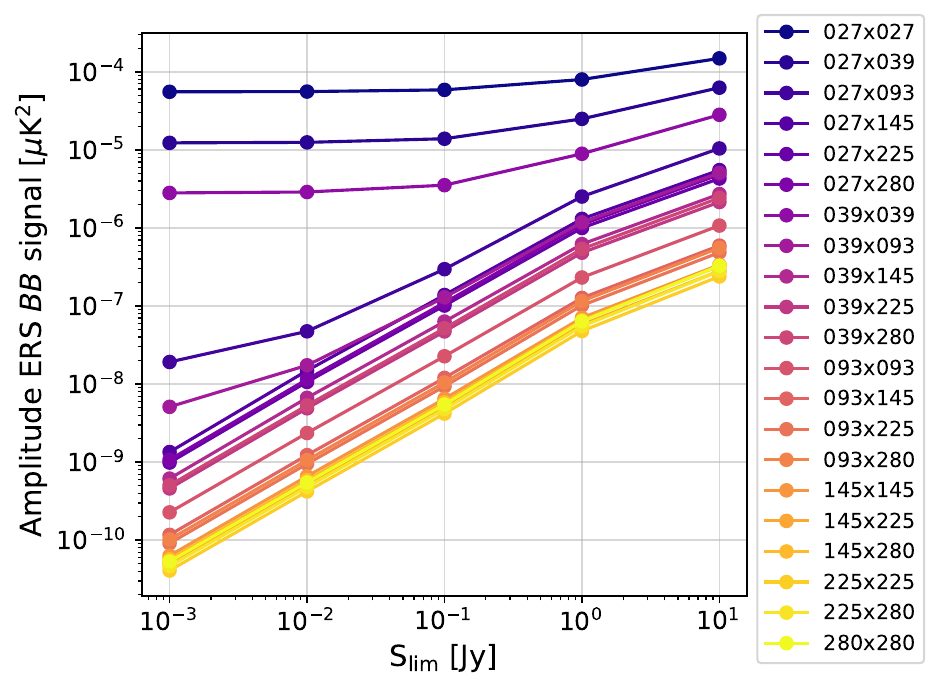}
        \caption{ERS frequency cross-spectral amplitudes after masking sources with fluxes $S> S_\mathrm{lim}$ at $\nu_m = 27\,{\rm GHz}$, with $f_{\rm FWHM}=1$, as a function of $S_\mathrm{lim}$.}
        \label{fig:amplitude_PS_as_function_of_Slim}
      \end{figure}

\begin{figure*}
        \begin{minipage}[b]{0.48\textwidth}
          \centering
          \includegraphics[width=\linewidth]{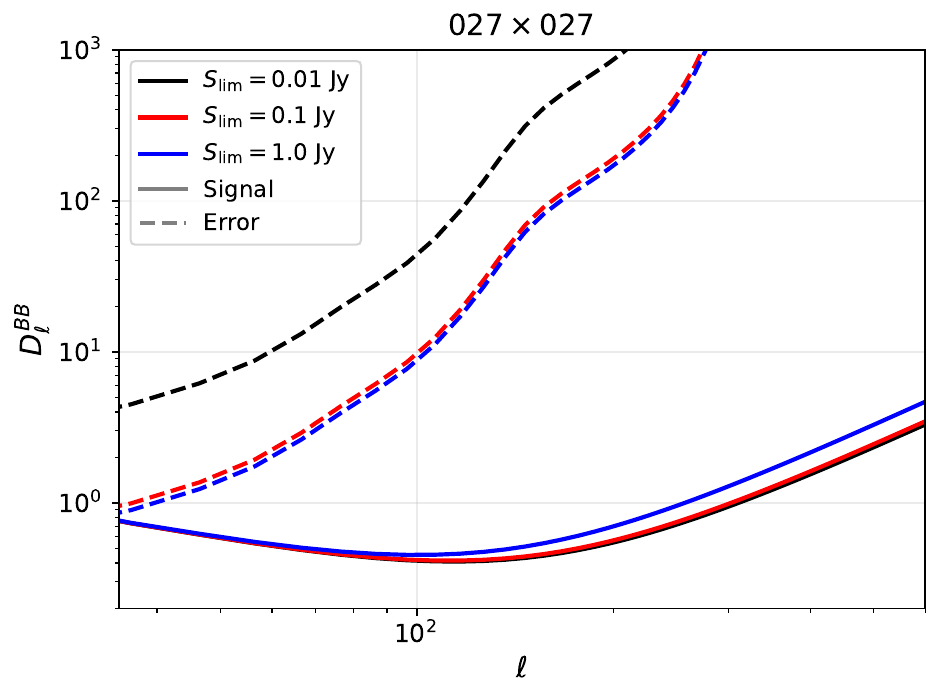}
        \end{minipage}
        \hspace{0.02\textwidth}
        \begin{minipage}[b]{0.48\textwidth}
          \centering
          \includegraphics[width=\linewidth]{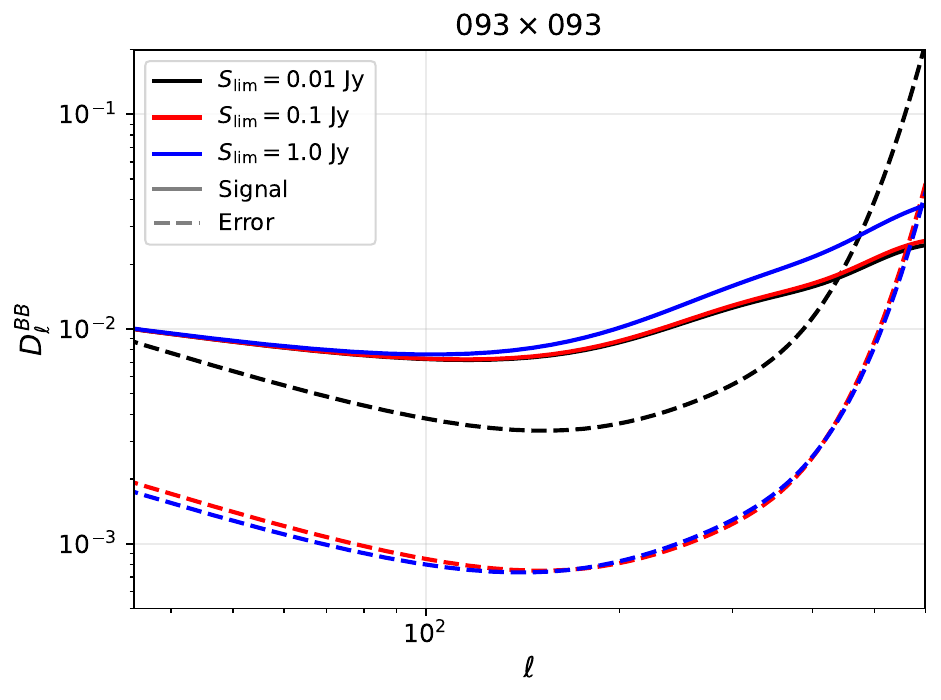}
        \end{minipage}
        \caption{For the extended configuration, we show an example of the power spectra obtained for the different flux cuts considered, focusing on the 27~GHz (\textit{left}) and 93~GHz (\textit{right}) bands. \textit{Signal} is the band-power values of the computed cross-spectra, while \textit{Error} is estimated as the square root of the diagonal elements of the covariance matrix. Note that the two plots do not share the same y-axis range. At 27~GHz, the ERS signal has a larger amplitude than in the other frequency bands, but is also affected by higher noise, particularly on small scales. At 93~GHz, even if the absolute ERS signal is lower than at 27~GHz, the combination of the lowest noise level among all frequency bands, the high angular resolution, and a relatively low Galactic foreground contribution results in the highest ERS signal-to-noise ratio. Comparing the different $S_\mathrm{lim}$ cases, the power spectrum uncertainties decrease as $S_\mathrm{lim}$ increases, thanks to the smaller masked sky area. At the same time, the residual ERS contamination grows, and can even dominate the signal on small scales. The latter effect can be appreciated in the different small-scale behaviour of the signal for the different $S_{\rm lim}$ values.
        }
        \label{fig:spectra_extended}
      \end{figure*}

      We show in Fig.~\ref{fig:amplitude_PS_as_function_of_Slim} residual contribution to multi-frequency power spectra from the residual ERS contamination as a function of $S_{\rm lim}$ for $f_{\rm FWHM}=1$. As noted in the previous section, the two lowest-frequency channels are the most affected by the residual flux of masked sources, manifested here in the plateau found for $S_{\rm lim}\lesssim0.1\,{\rm Jy}$ in the case of the auto- and cross-correlations of both of these channels. To better understand the impact of PS contamination on the data, Fig. \ref{fig:spectra_extended} shows the angular auto-power spectra at $\nu=27\,{\rm GHz}$ and $93\,{\rm GHz}$, including all contributions (CMB, Galactic foregrounds, and residual PS contamination) for three flux cuts: $S_{\rm lim}=1\,{\rm Jy}$, $0.1\,{\rm Jy}$, and $0.01\,{\rm Jy}$. The error power spectra for both frequencies is also shown for comparison. Owing to the better resolution and white-noise amplitude, the 93 GHz achieves a much better sensitivity across all frequencies, and is comparatively more strongly affected by the presence of point sources, particularly on scales $\ell\gtrsim150$. This reinforces the point made in the previous section, when choosing a value of $f_{\rm FWHM}$ that ensures point sources are fully masked at the more sensitive central channels. The amplitude of the spectra are visibly enhanced above $\ell\simeq150$ (more evidently so for the higher flux cut), following a $D_\ell\sim\ell^2$ behaviour characteristic of both lensing $B$ modes (at these scales) and point sources. On larger scales the power spectrum is dominated by Galactic foregrounds with a red scale dependence.
      \begin{figure*}
        \centering
        \includegraphics[width=\textwidth]{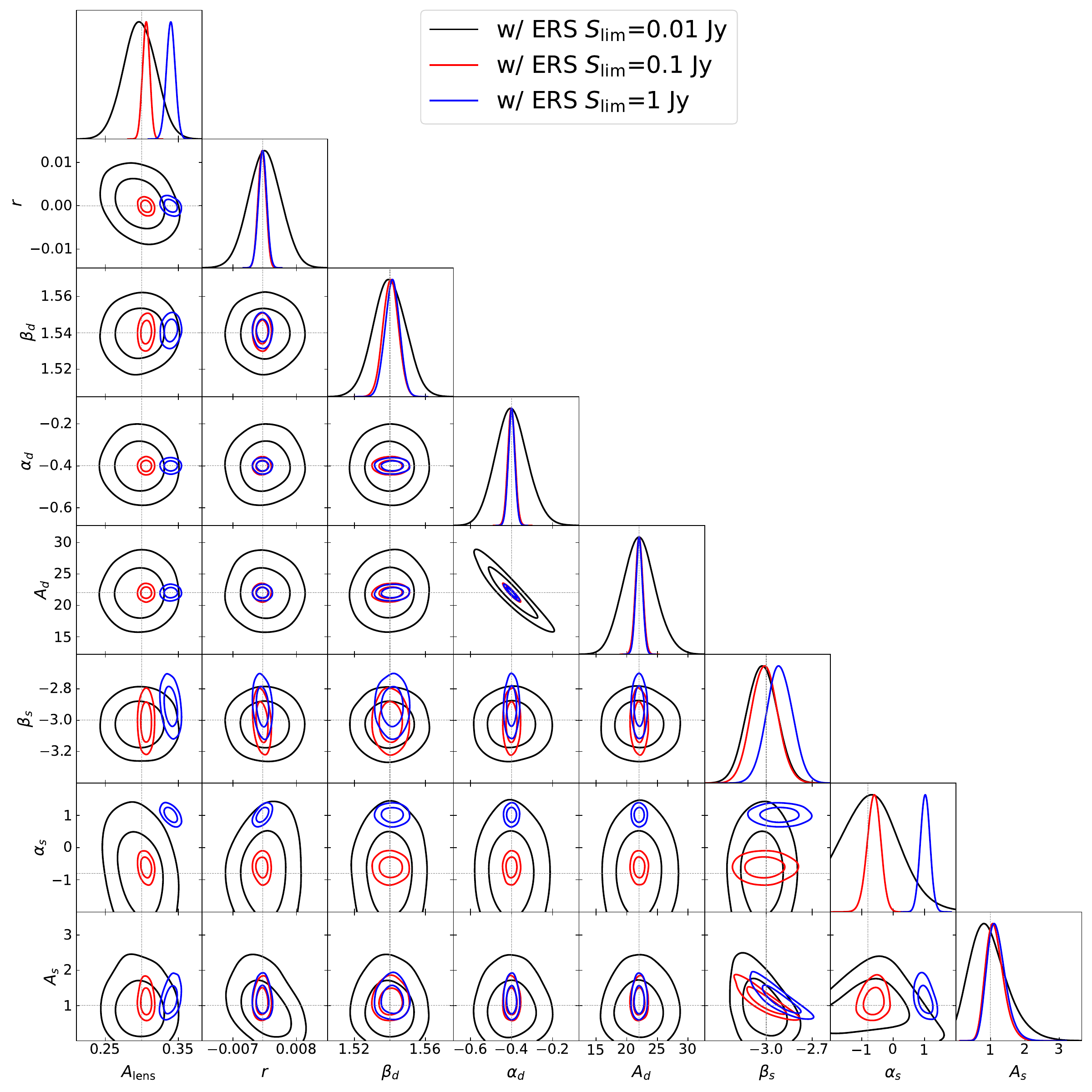}
        \caption{Constraints on cosmological and Galactic foreground parameters for the extended noise configuration, in the presence of unmodelled residual point source (ERS) contamination after masking sources with flux thresholds $0.01 \, \mathrm{Jy}$, $0.1 \, \mathrm{Jy}$ and $1 \, \mathrm{Jy}$. The multipole range used is $32 < \ell \leq 600$.}
        \label{fig:params_extended_0.01_0.1_1}
      \end{figure*}

      Figure \ref{fig:params_extended_0.01_0.1_1} shows the parameter constraints obtained for these flux limits, for the extended noise configuration. Comparing the cases $S_{\rm lim}=0.01\,{\rm Jy}$ and $0.1\,{\rm Jy}$ we find that the former leads to an excessive increase in statistical uncertainties, with the error on $r$ growing by $\sim 280\%$. 
      On the other hand, $S_\mathrm{lim} = 1 \, \mathrm{Jy}$ introduces very significant biases on the synchrotron and lensing parameters. In particular $\alpha_s$ is biased high by $12 \sigma$ ($2 \sigma$ for the baseline configuration), recovering a growing ``synchrotron'' spectrum (in fact corresponding to the PS contamination). 
      Interestingly, $\beta_s$ also experiences a milder $\sim1\sigma$ upwards bias, shifting the synchrotron SED to the larger spectral indices found in extragalactic sources on average. 
      Finally, $A_\mathrm{lens}$ is biased high by nearly $7 \sigma$ ($3 \sigma$ for baseline). 
      These large biases disappear almost completely when adopting a more conservative $S_{\rm lim}=0.1\,{\rm Jy}$. $A_{\rm lens}$ and $\alpha_s$ are now only $\sim1\sigma$ high ($\sim0.3\sigma$ for baseline). 
      In turn, the parameter errors grow only modestly: $20\%$ for $A_{\rm lens}$, $40\%$ for $\alpha_s$, and around $10\%$ or less for all the other parameters. We summarise the relevant information for these three cases and both configurations in Table~\ref{tab:results}. 
      The table also reports the $\chi^2$ values associated with the best-fit model in each case. Since the synthetic data was generated without statistic noise, these values should be understood as ``excess'' $\chi^2$ due to the presence of PS contamination. 
      In all cases explored in this section, even if the $\chi^2$ increases for larger values of $S_{\rm lim}$, the degradation to the goodness of fit is still rather small, and one would not suspect the presence of ERS contamination purely on the basis of this metric.
      \begin{table*}
        \centering
        \setlength{\tabcolsep}{3pt}
        \textbf{Baseline w/ nominal beams}
        \begin{tabular}{ccccccccccc}
        \hline
        $S_{\rm lim} [\mathrm{Jy}]$
        & $\mathrm{N}_{\rm PS}$
        & $f_{\rm sky} [\%]$
        & $r (\times10^3)$[true:0]
        & $A_{\rm lens}$[true:0.3]
        & $\alpha_s$[true:-0.8]
        & \Bs[true:-3]
        & $A_s$[true:$1$]
        & $\beta_{p}$
        & $A_{p} (\times 10^{-4})$
        & $\chi^2$
        \\
        \hline
        \hline
        \multicolumn{11}{c}{\rule{0pt}{2.5ex}ERS not modelled} \\
        \hline
        0.01 & 25144 & 0.4 & $-2.3 \pm 11.4$ & $0.282 \pm 0.077$ & $-0.02 \pm 0.98$ & $-3.03 \pm 0.10$ & $1.86 \pm 1.57$ & - & - & $0.03$ \\
        0.1 & 1699 & 7.8 & $0.1 \pm 2.9$ & $0.303 \pm 0.017$ & $-0.60 \pm 0.64$ & $-3.03 \pm 0.10$ & $1.16 \pm 0.57$ & - & - & $0.2$ \\
        1 & 88 & 9.6 & $-1.0 \pm 3.0$ & $0.352 \pm 0.020$ & $0.28 \pm 0.50$ & $-3.02 \pm 0.10$ & $1.91 \pm 0.70$ & - & - & $4$ \\
        \hline
        \multicolumn{11}{c}{\rule{0pt}{2.5ex}ERS modelled as PL} \\
        \hline
        0.01 & 25144   & 0.4 & $-3.4 \pm 11.7$ & $0.354 \pm 0.166$ & $0.38 \pm 1.07$ & $-3.03 \pm 0.10$ & $3.40 \pm 5.14$ & $-2.38 \pm 0.84$ & $-6.4 \pm 14.0$ & $0.09$ \\
        0.1 & 1699 & 7.8 & $0.0 \pm 3.1$ & $0.303 \pm 0.042$ & $-0.46 \pm 0.80$ & $-3.03 \pm 0.10$ & $1.27 \pm 0.80$ & $-2.28 \pm 0.87$ & $-0.3 \pm 1.7$ & $0.1$ \\
        1 & 88 & 9.6 & $0.0 \pm 2.9$ & $0.310 \pm 0.044$ & $-0.45 \pm 0.76$ & $-3.02 \pm 0.10$ & $1.28 \pm 0.69$ & $-2.33 \pm 0.78$ & $1.8 \pm 1.4$ & $0.16$ \\
        \hline
        \hline
        \end{tabular}
        \hspace{0.04\textwidth}
        \textbf{Extended w/ nominal beams}
        \begin{tabular}{ccccccccccc}
        \hline
        $S_{\rm lim} [\mathrm{Jy}]$ & $\mathrm{N}_{\rm PS}$ & $f_{\rm sky} [\%]$ & $r  (\times10^3)$[true:0] & $A_{\rm lens}$[true:0.3] & $\alpha_s$[true:-0.8] & \Bs[true:-3] & $A_s$[true:$1$] & $\beta_{p} $ & $ A_{p} (\times 10^{-4})$ & $\chi2$ \\
        \hline
        \hline
        \multicolumn{10}{c}{\rule{0pt}{2.5ex}ERS not modelled} \\
        \hline
        0.01 & 21544 & 0.4 & $0.5 \pm 3.8$ & $0.297 \pm 0.022$ & $-0.60 \pm 0.79$ & $-3.03 \pm 0.10$ & $0.99 \pm 0.54$ & - & - & $0.1$ \\
        0.1 & 1699 & 7.8 & $-0.1 \pm 0.9$ & $0.306 \pm 0.005$ & $-0.62 \pm 0.22$ & $-3.01 \pm 0.09$ & $1.13 \pm 0.26$ & - & - & $1$ \\
        1 & 88 & 9.6 & $0.0 \pm 1.0$ & $0.340 \pm 0.006$ & $1.02 \pm 0.15$ & $-2.91 \pm 0.08$ & $1.18 \pm 0.27$ & - & - & 50 \\
        \hline
        \multicolumn{11}{c}{\rule{0pt}{2.5ex}ERS modelled as PL} \\
        \hline
        0.01 & 25144 & 0.4 & $0.3 \pm 3.9$ & $0.304 \pm 0.041$ & $-0.27 \pm 0.95$ & $-3.02 \pm 0.10$ & $1.11 \pm 0.68$ & $-2.24 \pm 0.85$ & $-0.4 \pm 1.4$ & $0.08$ \\
        0.1 & 1699 & 7.8 & $0.0 \pm 0.9$ & $0.299 \pm 0.010$ & $-0.78 \pm 0.29$ & $-3.01 \pm 0.09$ & $1.06 \pm 0.27$ & $-2.20 \pm 0.83$ & $0.2 \pm 0.3$ & $0.2$ \\
        1 & 88 & 9.6 & $0.0 \pm 0.9$ & $0.301 \pm 0.015$ & $-0.78 \pm 0.31$ & $-3.02 \pm 0.09$ & $1.06 \pm 0.28$ & $-2.07 \pm 0.25$ & $2.3 \pm 0.3$ & $0.4$ \\
        \hline
        \hline
        \end{tabular}
        \hspace{0.04\textwidth}
        \textbf{Extended high resolution}
        \begin{tabular}{ccccccccccc}
        \hline
        $S_{\rm lim} [\mathrm{Jy}]$ & $\mathrm{N}_{\rm PS}$ & $f_{\rm sky} [\%]$ & $r (\times10^3)$[true:0] & $A_{\rm lens}$[true:0.3] & $\alpha_s$[true:-0.8] & \Bs[true:-3] & $A_s$[true:$1$] & $\beta_{p}$ & $A_{p}  (\times 10^{-4})$ & $\chi^2$ \\
        \hline
        \hline
        \multicolumn{11}{c}{\rule{0pt}{2.5ex}ERS not modelled} \\
        \hline
        0.01 & 21544 & 7.3& $-0.1 \pm 0.5$ & $0.301 \pm 0.003$ & $-0.72 \pm 0.04$ & $-3.00 \pm 0.02$ & $0.97 \pm 0.04$ & - & - & 5\\
        0.1 & 1699 & 9.5 & $-0.2 \pm 0.5$ & $0.308 \pm 0.003$ & $-0.16 \pm 0.04$ & $-2.91 \pm 0.02$ & $0.70 \pm 0.03$ & - & - & 237 \\
        1 & 88 & 9.7 & $-1.0 \pm 0.5$ & $0.328 \pm 0.003$ & $1.41 \pm 0.03$ & $-2.36 \pm 0.01$ & $0.21 \pm 0.01$ & - & - & 772 \\
        \hline
        \multicolumn{11}{c}{\rule{0pt}{2.5ex}ERS modelled as PL}\\
        \hline
        0.01 & 25144 & 7.3 & $0.0 \pm 0.6$ & $0.301 \pm 0.003$ & $-0.79 \pm 0.05$ & $-3.00 \pm 0.02$ & $1.00 \pm 0.04$ & $-2.87 \pm 0.60$ & $0.0014 \pm 0.0007$ & $0.4$ \\
        0.1 & 1699 & 9.5 & $-0.1 \pm 0.5$ & $0.303 \pm 0.003$ & $-0.79 \pm 0.05$ & $-3.00 \pm 0.02$ & $1.00 \pm 0.03$ & $-2.38 \pm 0.06$ & $0.0093 \pm 0.0007$ & 1 \\
        1 & 88 & 9.7 & $-0.1 \pm 0.5$ & $0.319 \pm 0.004$ & $-0.73 \pm 0.07$ & $-3.01 \pm 0.02$ & $0.99 \pm 0.04$ & $-2.22 \pm 0.01$ & $0.0649 \pm 0.0013$ & 28 \\
        \hline
        \end{tabular}
        \caption{Summary of the mean values and 1$\sigma$ confidence interval from the marginalised posteriors for the most relevant parameters for the different configurations studied. Galactic dust parameters remain unbiased in all cases explored and are not shown. The amplitude parameters $A_s$ and $A_p$ are in units of $\mu \mathrm{K}^2$.}\label{tab:results}
      \end{table*}

    \subsubsection{Marginalising over residual PS contamination}\label{sssec:res.ers.marg}
      Although a PS flux limit of $S_\mathrm{lim}=0.1 \, \mathrm{Jy}$ provides a reasonable balance between bias and variance, a small $\sim1\sigma$ bias in $A_\mathrm{lens}$ and $\alpha_s$ remains, which could affect the interpretation of $B$-mode analyses with future datasets. In addition to this, current uncertainties in the modelling of polarised point sources at $20$-$200\,{\rm GHz}$ frequencies are significant, and could make this mild biases more significant. It is therefore interesting to explore whether the residual PS contamination may be effectively marginalised over, and what the penalty of doing so is in terms of posterior parameter constraints.

      To explore this, we introduce a fourth sky component incorporating this contamination. We model the effective SED of this component as a power law in Rayleigh-Jeans unit, with a spectral index $\beta_p$ ($f_p(\nu)=(\nu/\nu_0^p)^{\beta_p}$), and with a flat/white power spectrum, characterised by an amplitude $A_p$. In detail, this can only be an approximate model for the residual contamination, not only because the combination of emission from different ERS samples is not described by an exact power law, but also because the distinct effective masking of sources at different frequencies will cause further frequency decorrelation. Nevertheless, this simplistic model may be enough to capture the bulk of the residual contamination, reducing parameter biases below their statistical uncertainties.

\begin{figure*}
        \centering
        \includegraphics[width=\textwidth]{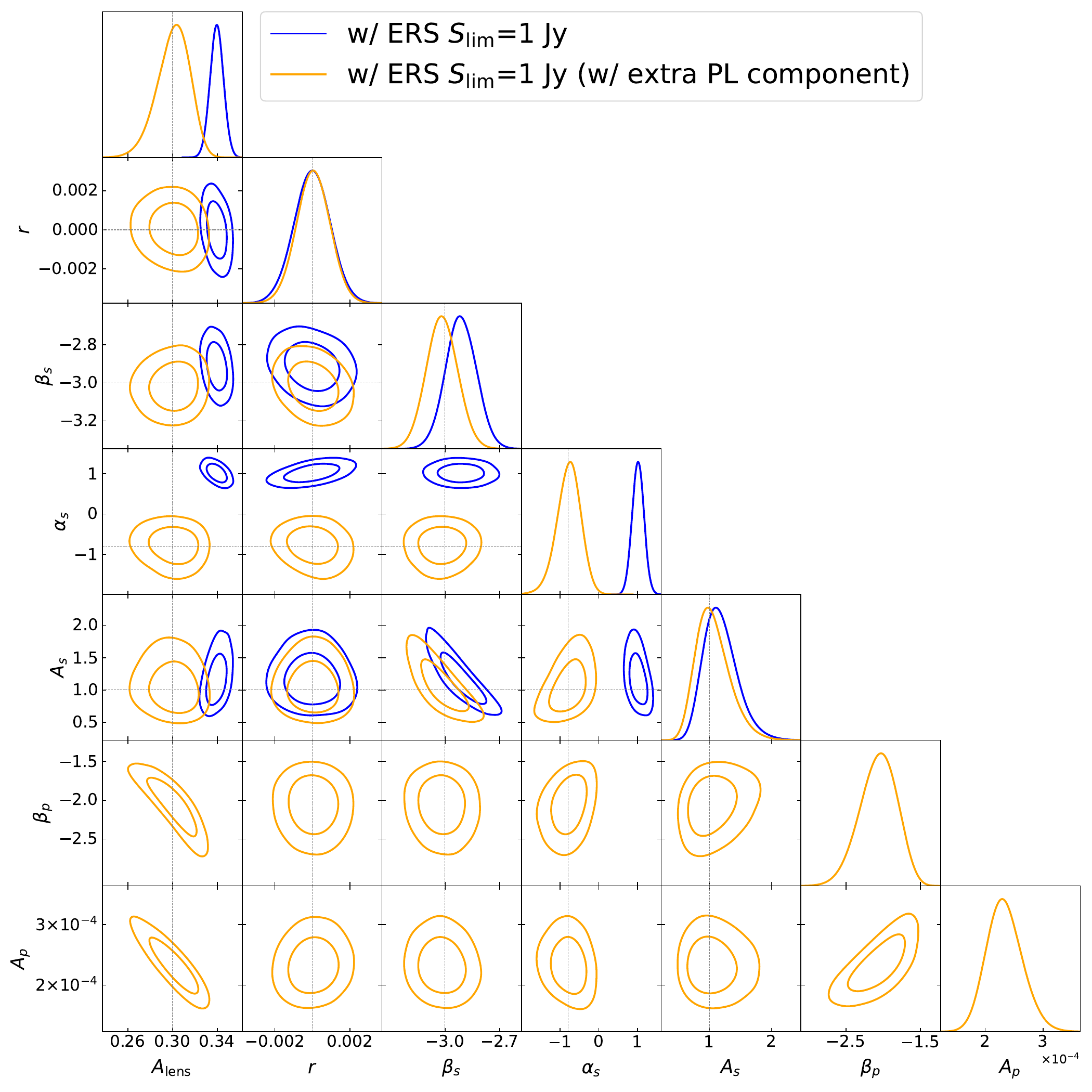}
        \caption{As Fig. \ref{fig:params_extended_0.01_0.1_1} for a flux limit $S_{\rm lim}=1 \, \mathrm{Jy}$, comparing the cases where the residual point source contamination is left unmodelled (blue) or modelled as an additional component with a power-law SED with spectral index $\beta_p$ and a flat power spectrum with amplitude $A_p$.}\label{fig:parameters_extended_Slim_1.0_w_PS_modelling}
      \end{figure*}

\begin{figure}
        \centering
        \includegraphics[width=\linewidth]{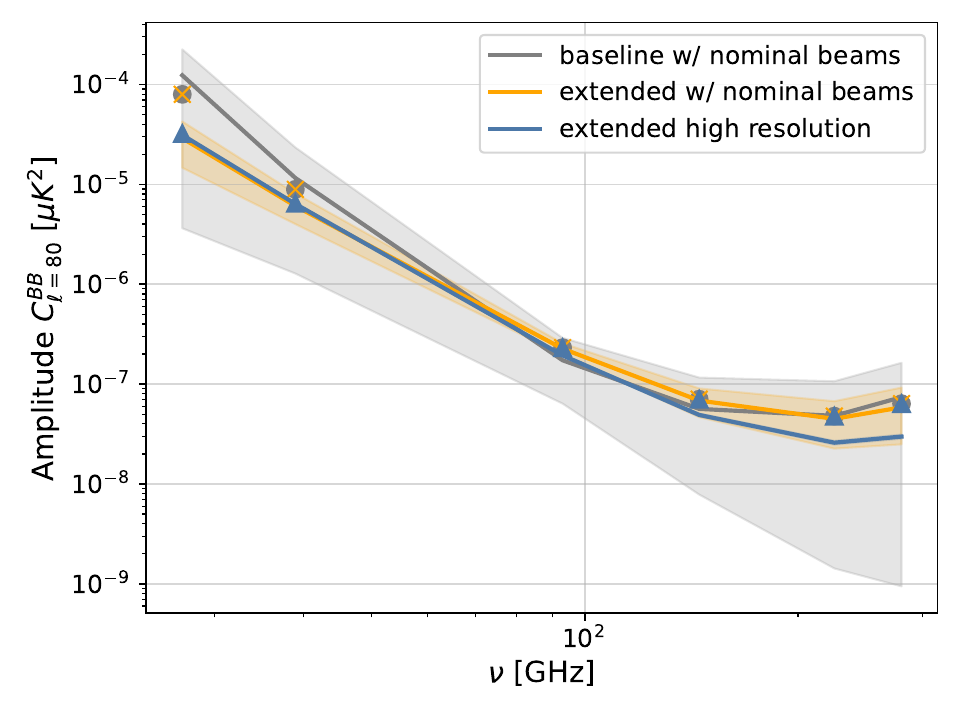}
        \caption{Reconstructed ERS power spectral amplitude (lines) compared with the true input ERS signal (markers), for the $S_\mathrm{lim}=1 \, \mathrm{Jy}$ case. In the cases with nominal beams the pivot frequency is $\nu_0 = 93 \, \mathrm{GHz}$, which explains the deviations at high and low frequencies. For the high-resolution case we use $\nu_0 = 23 \, \mathrm{GHz}$, and the low-frequency ERS signal is well recovered, while the model deviates more at high frequencies. The shaded areas corresponds to the $68 \%$ confidence interval.}
        \label{fig:point_sources_true_vs_reconstructed_Slim_1.0_extended}
      \end{figure}

      The result of adding this component on the inferred model parameters is shown in Table \ref{tab:results} for the different configurations explored here, and in Fig. \ref{fig:parameters_extended_Slim_1.0_w_PS_modelling} in the case $S_{\rm lim}=1\,{\rm Jy}$ for the extended noise configuration. Overall we find that including the additional component resolves the parameter biases described in the previous section in almost all cases, even those that exhibit a significant residual ERS signal. For instance, where before we found biases of up to $7\sigma$ in $A_s$, and $12\sigma$ in $\alpha_s$ for $S_{\rm lim}=1\,{\rm Jy}$, both parameters are recovered with no apparent bias when including the additional component. A penalty is paid in the form of a $\sim2$-fold increase in the posterior uncertainties in both parameters, although the constraints on $r$ remain largely unchanged. From the constraints on the amplitude $A_p$, we find that the PS residual is detected at the $\sim7.5\sigma$ level. In cases where the residual contamination is less significant, either due to using a more stringent flux cut, or to the larger noise in the baseline configuration, a mild bias in $\alpha_s$, below the $1\sigma$ level, remains. This is largely due to the three-way degeneracy between $A_p$, $\beta_p$, and $\alpha_s$, leading to volume/projection effects in the posterior constraints. All other parameters, including $A_{\rm lens}$, remain unbiased, however. An important point in this is context is the choice of pivot frequency $\nu_0^p$ for this component: when the residual PS contribution is close to the threshold of detection, fitting it with a non-linear two-parameter model ($A_p$, $\alpha_p$) can lead to strong projection effects when $\nu_0^p$ is far from the channel with the highest SNR. Although these may be avoided by imposing a Jeffreys prior on the non-linear parameter $\alpha_p$, we find that these effects largely disappear when we choose $\nu_0^p=93\,{\rm GHz}$, our most sensitive channel. With this choice, we show in Fig. \ref{fig:point_sources_true_vs_reconstructed_Slim_1.0_extended} that we are able to accurately reconstruct the true residual ERS signal amplitude at 93~GHz, at the price of measuring a slightly biased amplitude at the lower frequencies. In other words, setting the pivot frequency at 93~GHz we can correctly measure $A_p$, but we can not constrain $\alpha_p$ when there is insufficient signal at the other frequencies.

      In summary, we conclude that this simple effective model (a Poisson-like component with power-law SED) is sufficient to capture the residual point source contamination, even with minimal masking ($S_{\rm lim}=1\,{\rm Jy}$), and recover unbiased constraints on cosmological and foreground parameters, without the need to fit for individual white noise amplitudes in each cross-frequency spectrum.

    \subsubsection{High angular resolution}\label{sssec:res.ers.hires}
      The impact of point sources in the cases studied so far is mitigated by the fact that the ERS signal, peaking at low frequencies, dominates over DSFGs (see Section \ref{ssec:res.dsfg}). 
      At these low frequencies, SAT-like instruments have poor sensitivity on small scales, where PS contamination may dominate over Galactic foregrounds. It is therefore interesting to quantify the impact of ERS contamination on potential future experiments with better angular resolution.

      To do so we study an ideal case in which the instrument beam does not impact observations on the angular scales considered here, $32 < \ell \leq 600$. Specifically, we repeat our analysis assuming all frequency bands to be characterised by the same Gaussian beam with $\theta_{\rm FWHM} = 9'$. In this case we are able to mask much more of each PS without incurring a large penalty in terms of area loss, even when masking $\sim10^4$ sources. Therefore we use $f_\mathrm{FWHM}=3$ (i.e. fully masking each source at all frequencies) for this analysis. Furthermore, the fact that, in this high-resolution setup, all the bands have the same beam, means that the astrophysical frequency scaling of the point sources is now preserved.

      The results of this study are summarised in Table \ref{tab:results}. We find that, unsurprisingly, the improved sensitivity on small scales increases all parameter biases due to residual contamination. In particular $A_{\rm lens}$ remains biased at the $\gtrsim3\sigma$ level even for $S_{\rm lim}=0.1\,{\rm Jy}$, where all synchrotron parameters also exhibit large biases. However, a more stringent flux cut, at $S_{\rm lim}=0.01\,{\rm Jy}$, suppresses these biases almost entirely, at no significant cost in the posterior parameter uncertainties. The only exception is the synchrotron tilt $\alpha_s$,  which retains a $\sim2\sigma$ bias. $r$, as well as all dust parameters, remain unbiased in almost all cases, although the least stringent setup, with $S_{\rm lim}=1\,{\rm Jy}$ recovers a $\sim2\sigma$ negative bias on $r$.

      \begin{figure}
        \centering
        \includegraphics[width=\linewidth]{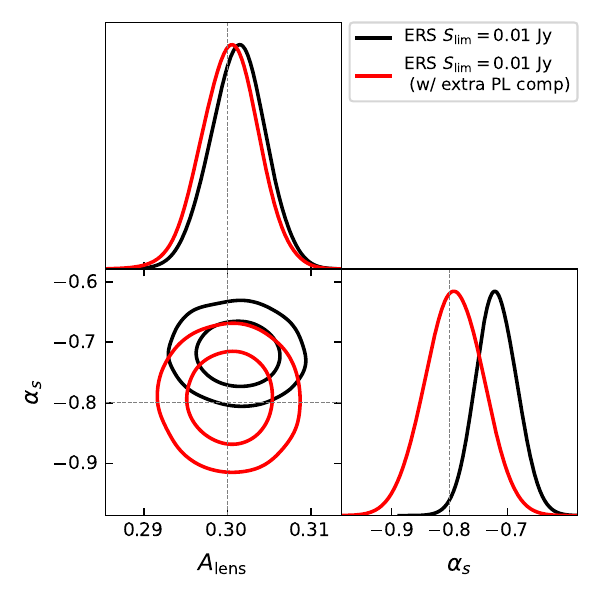}
        \caption{Constraints on the lensing $B$-mode amplitude $A_{\rm lens}$ and the synchrotron spectral tilt $\alpha_s$ in the high-resolution instrumental configuration for the $S_{\rm lim}=0.01\,{\rm Jy}$ case. We compare the cases where the residual PS contamination is left unmodelled (black), or marginalised over using an additional component with an effective power law SED and a white power spectrum (red). The multipole range used is $32 < \ell \leq 600 $.}\label{fig:params_high_reso_Slim0.01}
      \end{figure}
      These results suggest that, although the use of a more stringent mask may be sufficient to reduce the most important sources of bias, we must inevitably marginalise over residual contamination to fully recover unbiased synchrotron constraints. We thus repeat our analysis including a fourth effective component, as described in the previous section, characterised by a spectral index $\beta_p$, and a Poisson amplitude $A_p$. The results are summarised in Table \ref{tab:results}. We find that including this additional component cancels the residual bias on $\alpha_s$ in the case with $S_{\rm lim}=0.01\,{\rm Jy}$ (see also Fig. \ref{fig:params_high_reso_Slim0.01}), and suppresses almost all parameter biases for $S_{\rm lim}=0.1\,{\rm Jy}$ and even $1\,{\rm Jy}$, with the only exception of $A_{\rm lens}$ in the latter case, which retains a $\sim4.5\sigma$ bias. We did not find simple extensions to the effective point source models, such as including an effective frequency decorrelation parameter to characterise deviations from a perfect power law SED, to help reduce this bias. High-resolution experiments will therefore require a minimal level of polarised point source masking to achieve unbiased constraints.

    \subsubsection{Polarisation fraction scatter}\label{sssec:res.ers.polfrac}
      The largest source of uncertainty in our model for the ERS contamination is the assumed polarisation fraction, which is not well constrained by current data. Throughout our results we have assumed a constant polarisation fraction across the sources of $2.8 \%$ as motivated in \cite{lagache2020impact}. We now explore a more conservative scenario in which the scatter in the distribution of polarisation fractions, as suggested by \cite{puglisi2018forecasting}, increases the effective fraction relevant to the residual ERS power spectrum to $\langle \Pi^2 \rangle^{1/2} =4.94\%$. This is the effective fraction, accounting for scatter, estimated for flat spectrum sources in \cite{puglisi2018forecasting}, which dominate the ERS emission at the frequencies studied here. This change corresponds to rescaling the ERS cross-spectra we used so far by a factor 3.1.

      Unsurprisingly, this enhances the biases found in previous sections by a similar amount. For example, for $S_\mathrm{lim}=0.1 \, \mathrm{Jy}$, the biases on $A_{\rm lens}$ and $\alpha_s$ grow from $\sim1\sigma$ to $\sim3\sigma$. The tensor-to-scalar ratio, as well as all dust parameters, remain unbiased. However, as before, all these biases disappear when including an effective Poisson-like component with a power-law SED to describe the residual ERS contamination. In this case, this effective component is detected at a higher (although still marginal) significance of $2.3 \sigma$.

  \subsection{Dusty galaxies}\label{ssec:res.dsfg}
    \begin{figure}
        \centering
        \includegraphics[width=\linewidth]{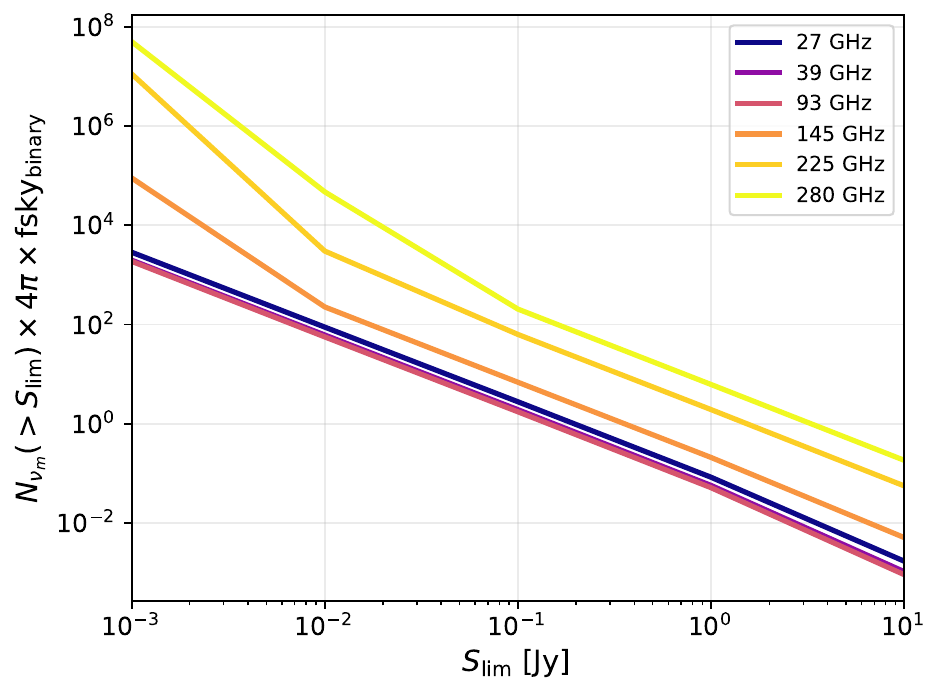}
        \caption{Total DSFG number counts in the sky patch shown in Fig. \ref{fig:mask} for different $\nu_m$ as a function of $S_\mathrm{lim}$.}
        \label{fig:num_counts_DSFG_as_function_of_Slim_and_nu_mask}
    \end{figure}
    \begin{figure*}
      \centering
      \includegraphics[width=\linewidth]{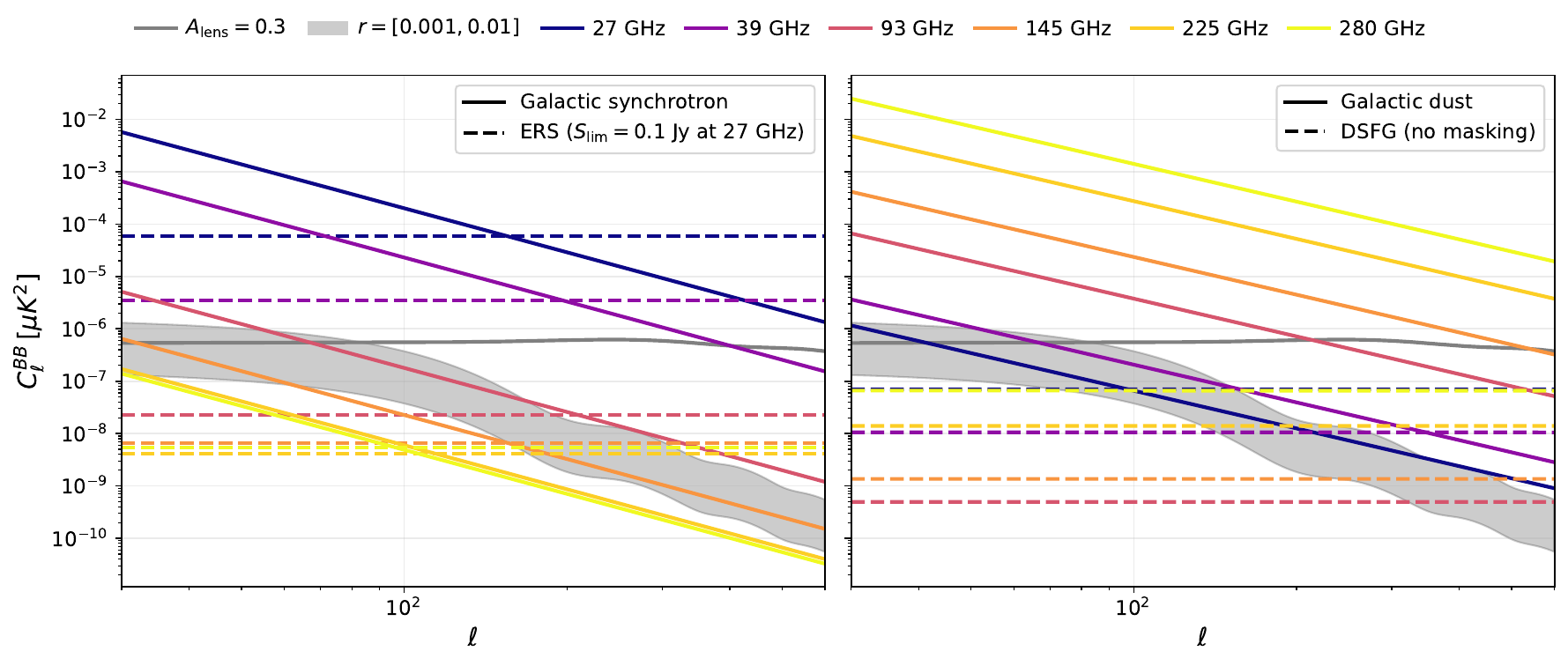}
      \caption{\textit{Left}: power spectra for the Galactic synchrotron signal and the ERS signal with $S_\mathrm{lim} = 0.1 \, \mathrm{Jy}$ and masking performed at 27~GHz as described in the text. \textit{Right}: power spectra for the Galactic dust signal and the full DSFG signal (no masking). Both are shown in the multipole range relevant for our analysis, and the CMB signal is over-plotted (lensing $B$ modes with $A_\mathrm{lens}=0.3$ and primordial $B$ modes in the range $0.001<r<0.01$, relevant for upcoming CMB experiments.)} \label{fig:xcls_synch_ERS_dust_DSFG}
    \end{figure*}
    We now consider the impact of polarised DSFGs, modelled as described in Sect.~\ref{ssec:model.DSFG}. 
    First, we compute the number counts for the DSFGs, see Fig.~\ref{fig:num_counts_DSFG_as_function_of_Slim_and_nu_mask}.
    Contrary than what found for the ERS, for a fixed $S_\mathrm{lim}$ the number of sources decreases towards lower frequency, hence favouring masking at $\nu_m=280 \, \mathrm{GHz}$.
    And the number count differences among the different frequencies is more important for DSFGs than ERSs.
    Also, comparing with Fig.~\ref{fig:num_counts_ERS_as_function_of_Slim_and_nu_mask}, we find that there are more faint DSFGs than ERSs, but less bright DSFGs than ERSs. 
    In particular, in the SO SAT sky area, at $S_\mathrm{lim} \simeq 1 \, \mathrm{Jy}$ there are only $\sim 6$ DSFGs at $\nu_m=280 \, \mathrm{GHz}$, while $\sim 88$ DSFGs at $\nu_m=27 \, \mathrm{GHz}$.
    Such a low number of DSFGs above $S_\mathrm{lim} \simeq 1 \, \mathrm{Jy}$ hints that we may avoid masking DSFGs at all.
        
    Then, it is instructive to compare the ERS and DSFGs signals, both with each other and with Galactic foregrounds. Fig.~\ref{fig:xcls_synch_ERS_dust_DSFG} shows, in the left panel, the residual ERS power spectrum for a flux limit $S_{\rm lim}=0.1\,{\rm Jy}$ together with the Galactic synchrotron signal in the SAT patch. In turn, the right panel shows the estimated polarised DSFG signal without masking any sources, together with the Galactic dust component. All signals are shown for the auto-correlation of the six frequency channels studied here. The figure also shows the lensing signal for $A_{\rm lens}=0.3$, and the primordial $B$-mode signal for $r$ in the range $0.001\leq r\leq0.01$ for comparison.

    Already at the power spectrum level we can see that DSFGs are subdominant with respect to ERSs. At every frequency band, except 225~GHz and 280~GHz, the $BB$ power spectra of the DSFGs is lower than the ERS signal, even when we mask out ERSs with $S_{\rm lim}=0.1\,{\rm Jy}$ and do not mask any DSFGs. In addition to this, the Galactic dust signal is higher than Galactic synchrotron, and dominates over the DSFG contribution at almost all scales relevant for this analysis. Finally, the DSFG signal is lower than the $B$-mode lensing contribution at all relevant scales. 

    It is worth noting that the DSFG signal, as modelled here, reaches a minimum at around $68\,{\rm GHz}$, growing towards higher frequencies with a dust-like modified black-body spectrum, but also towards lower frequencies, due to the synchrotron contribution from star-forming galaxies. Thus, the signal at 27 and 39 GHz is comparable to that in the higher-frequency channels. The larger beam limits the sensitivity to this contribution at the lower frequencies, however.

    Finally, it is worth recalling that the DSFG signal discussed here contains both a shot-noise and clustered 1-halo contribution, although the former dominates over the latter here.

    We propagate the DSFG residual shown in Fig. \ref{fig:xcls_synch_ERS_dust_DSFG} to parameter constraints as done before in the case of ERSs. We find that, even without masking any DSFGs, as in the spectra shown in the figure, their contribution is negligible for both the baseline and extended configurations, with no impact on the recovery of either the Galactic foreground or the cosmological parameters. No significant bias is found on any of the model parameters, including $\alpha_s$, $\alpha_d$ or $A_{\rm lens}$. Even in the high-resolution configuration explored in Section \ref{sssec:res.ers.hires}, only a minor $0.3\sigma$ bias develops for $\alpha_s$, sourced by the radio component of the SED template used to describe DSFGs.

\section{Conclusions}\label{sec:conclusion}
  In this work we have studied the impact from polarised extragalactic point sources in ground-based $B$-mode analyses. Using the models of \cite{tucci2011high,toffolatti1998extragalactic,lagache2020impact} for ERSs and \cite{zagatti2024halo} for DSFGs, we have propagated the residual point source contamination through component separation into constraints on $r$, $A_{\rm lens}$, and Galactic foreground parameters. We have done so using a power-spectrum-based component separation pipeline, assuming a self-consistent masking strategy that uses the same point source mask at all frequencies. 

  We find that, while residual DSFG contamination does not lead to significant biases, ERS contamination must be mitigated and taken into account in near-future experiments, such as the Simons Observatory. Specifically focusing on small-aperture experiments, such as the SO SATs, we have shown that an effective strategy requires masking the bulk of the source contamination, without necessarily removing it entirely when the angular resolution varies substantially across frequency channels. Specifically, a flux cut of approximately $S_{\rm lim}=0.1\,{\rm Jy}$ at 27 GHz provides a suitable compromise between minimising bias and variance in the model parameters (although the exact value of $S_{\rm lim}$ may be optimised for individual instrumental configurations). More concretely, we find that, while the tensor-to-scalar ratio $r$ is largely unaffected by point source contamination, the lensing amplitude $A_{\rm lens}$ and the spectral tilt for Galactic synchrotron $\alpha_s$ may be significantly biased. We also show that any remaining parameter bias may be effectively suppressed by modelling the residual point source contamination as an additional component, with a power-law SED and a flat, Poisson-like, angular power spectrum, after marginalising over the free parameters of this model. Furthermore this is achieved at no cost in the statistical uncertainties on $r$, although the errors on $A_{\rm lens}$ and $\alpha_s$ approximately double.

  We additionally investigate an hypothetical experiment with higher angular resolution, where point source masking results in a much milder loss of survey area. In this case, a more stringent flux cut, $S_\mathrm{lim}=0.01\,\mathrm{Jy}$, achievable by high-resolution telescopes such as the SO LAT, provides the most effective mitigation strategy. This must be accompanied by the modelling and marginalisation over the residual contamination as described above.
  
  Our analysis has focused only on the impact of point source contamination at the power spectrum level, and we have not explicitly considered the practical implications of introducing a large number of source holes in the analysis mask, beyond the increase in power spectrum uncertainties due to the loss of survey area. In practice, utmost care must be taken in the presence of complex masks to minimise the level of $E$-to-$B$ leakage, which can severely degrade the precision of the $BB$ power spectrum. Such effects will, however, need to be carefully accounted for in applications to real data, where extensive point source masking may be unavoidable. A natural extension of the approach proposed here would therefore be to combine point source masking with inpainting techniques, in order to mitigate these effects. Finally, future, more sensitive experiments, or those covering larger sky areas (such as LiteBIRD) may be more affected by point source contamination than the configurations explored here, and may require the use of more strict masking criteria, or the implementation of more sophisticated model to describe the residual point source contributions than the relatively simple effective power law behaviour proposed here.

\section*{Acknowledgements}
  AR thanks Gilles Weymann-Despres for stimulating scientific discussions and acknowledges support from the Dennis Sciama Junior Research Fellowship at Wolfson College, Oxford.
  GZ is funded by the European Union (ERC, RELiCS, project number 101116027). GZ acknowledges the financial support from the COSMOS network (www.cosmosnet.it) through the ASI (Italian Space Agency) Grants 2016-24-H.0 and 2016-24-H.1-2018. 
  
  Note that this is not an official Simons Observatory collaboration publication.

\section*{Data Availability}

No new data were generated or analysed in support of this research.



\bibliographystyle{mnras}
\bibliography{bib}



\appendix

\bsp	
\label{lastpage}
\end{document}